\documentclass[aps,pra,twocolumn,floatfix,nofootinbib]{revtex4-2}
\usepackage[colorlinks]{hyperref}
\usepackage[utf8]{inputenc}
\usepackage{graphicx}
\usepackage{multirow}
\usepackage{amssymb}
\usepackage{amsmath}
\usepackage{microtype}
\usepackage{braket}
\usepackage{xcolor}

\def\rmel#1#2#3{ \langle #1|| #2 || #3 \rangle }
\def\mr{\mathrm}
\def\mb{\mathbf}
\def\bs{\boldsymbol}

\newcommand{\Cs}{\ensuremath{^{133}\mathrm{Cs}}}

\newcommand{\im}{{\rm i}}

\begin{document}
	\title{Parity-mixed coupled-cluster formalism for computing  parity-violating amplitudes}
	\author{H. B. Tran Tan}
	\author{Di Xiao}
	\author{A. Derevianko}
	\affiliation{Department of Physics, University of Nevada, Reno, Nevada 89557, USA}
\begin{abstract}
We formulate a parity-mixed coupled-cluster (PM-CC) approach for high-precision calculations of parity non-conserving amplitudes in mono-valent atoms. Compared to the conventional formalism which uses parity-proper (PP) one-electron orbitals, the PM-CC method is built using parity-mixed (PM) orbitals. The PM orbitals are obtained by solving the Dirac-Hartree-Fock equation with the electron-nucleus electroweak interaction included (PM-DHF). There are several advantages to such a PM-CC formulation: (i) reduced role of correlations, as for the most experimentally-accurate to date \Cs\ $6S_{1/2}-7S_{1/2}$ transition, the PM-DHF result is only 3\% away from the accurate many-body value, while the conventional DHF result is off by 18\%;  (ii) avoidance of directly summing over intermediate states in expressions for parity non-conserving amplitudes which reduces theoretical uncertainties associated with highly-excited and core-excited intermediate states, and (iii) relatively straightforward upgrade of existing and well-tested large-scale PP-CC codes. We reformulate the CC method in terms of the PM-DHF basis and demonstrate that the cluster amplitudes are complex numbers with opposite parity real and imaginary parts. We then use this fact to map out a strategy through which the new PM-CC scheme may be implemented.
\end{abstract}
	
\maketitle

\section{Introduction}\label{Introduction}
The field of parity violation started with the seminal paper by Lee and Yang \cite{Lee1956} and the discovery of parity non-conservation (PNC) in  nuclear $\beta$-decay \cite{Wu1957}. Shortly after, the possibility of measuring atomic parity violation (APV) as a low-energy test for the Standard Model (SM) was investigated by Zel'dovich \cite{Zeldovich1959}, whose consideration for hydrogen suggested that the effects were  too small to be observable. The situation changed when the Bouchiats \cite{BOUCHIAT1974_1,BOUCHIAT1974_2,BOUCHIAT1975} demonstrated that APV effects scale as $Z^3$, where $Z$ is the nuclear charge, thus reopening the case for observing them in heavy neutral atoms. Following a proposal by Khriplovich \cite{Khriplovich1974}, APV effects were first observed in bismuth by Barkov and Zolotorev \cite{Barkov1978}. Following this discovery, several APV experiments were performed for cesium \cite{BOUCHIAT1982,Wood1997,Guena2003,Guena2005,Tsigutkin2009}, bismuth \cite{Macpherson1991}, lead \cite{Meekhof1993,Phipp1996}, thallium \cite{Vetter1995,Edwards1995} and ytterbium \cite{antypas2019}. New APV experiments are underway or in the planning stage \cite{Gomez2005,DeMille2008,Antypas2013,choi2018gain,aubin2013atomic,portela2013towards,Altuntas2018,Aluntas2018_2} (see also the review~\cite{Safronova2018} and references therein), with the aim of attaining a $\sim 0.1\%$ accuracy in $^{133}$Cs.

The APV measurements are usually interpreted in terms of the nuclear weak charge $Q_W$, which is related to the measured PNC amplitude, $E_{\rm PV}$, via $E_{\rm PV}=k_{\rm PV}Q_W$, where $k_{\rm PV}$ is an atomic-structure factor. One wishes to compare the experimentally obtained value of $Q_W$ with the value predicted by the SM. For this purpose, the quantity $k_{\rm PV}$ should be known with a better accuracy than that of the amplitude $E_{\rm PV}$, thus yielding an accurate estimate of $Q_W$. This approach has been so far most successful in $^{133}$Cs, due to its large nuclear charge, $Z=55$, and its relatively simple atomic structure with one valence electron above a closed Xe-like core~\cite{WiemanDerevianko-SM50-2019}. Part of the success was also due to the fact that $^{133}$Cs is used in the primary frequency standard, providing a wealth of information on its basic atomic properties.

In \Cs, the experimental uncertainty for the $6S_{1/2}-7S_{1/2}$ PNC amplitude eventually reached 0.35\% \cite{Wood1997}. The most accurate theoretical computations for the atomic-structure factor $k_{\rm PV}$ was built upon the relativistic Many-Body Perturbation Theory (MBPT), a systematic order-by-order approach which includes electron correlations. Certain classes of MBPT diagrams can be summed to all orders, taking advantage of the underlying topology of the diagrams. In the late 1980s and early 1990s, the accuracy of MBPT calculations for $k_{\rm PV}$ was estimated to be at the level of 1\% \cite{Dzuba1985,DZUBA1989,Blundell1990,Blundell1992}. 

A later re-analysis~\cite{Bennett1999},  based on an improved theory-experiment agreement with the new measurements of atomic properties, reduced the theoretical uncertainty of Refs.~\cite{Dzuba1985,DZUBA1989,Blundell1990,Blundell1992} to the level of 0.4\%. The deduced value for the \Cs\ weak charge  differed by $2.5\sigma$ from the SM prediction, thus suggesting new physics beyond the SM \cite{Ramsey-Musolf1999,CASALBUONI1999,Rosner2000,Rosner2002}. However, the inclusion of Breit \cite{Derevianko2000,Derevianko2001,Dzuba2001} and QED radiative corrections \cite{Johnson2001,Dzuba2002,Shabaev2005,Flambaum2005} brought the \Cs\ result back into the essential $1\sigma$ agreement with the SM. Clearly, the answer to whether the \Cs\ PNC result confirms the SM or hints at new physics very much depends on the quality of theoretical atomic calculation for the atomic-structure factor $k_{\rm PV}$. In the works \cite{Derevianko2000,Derevianko2001,Dzuba2001,Johnson2001,Dzuba2002,Flambaum2005}, the theoretical uncertainty stood at $0.5\%$, still larger than the $0.35\%$ experimental error bar.

Since the early 2000s, the theoretical error bar has been (and still is) dominated by the uncertainty of solving the basic many-body problem of atomic structure. Further progress in improving the theoretical accuracy was reported in the late 2000s \cite{Porsev2009,Porsev2010}. These calculations built upon the \textit{ab initio} relativistic coupled-cluster (CC) scheme \cite{Blundell1990}. While the calculations of Refs.~\cite{DZUBA1989,Blundell1990,Blundell1992} were complete through the third order of MBPT for matrix elements, the scheme in Refs.~\cite{Porsev2009,Porsev2010} was complete through the fourth order of MBPT. References~\cite{Porsev2009,Porsev2010} have reduced the theoretical uncertainty in the \Cs\ atomic-structure factor $k_{\rm PV}$ to 0.27\%. The final value of the \Cs\ weak charge extracted from this calculation was in an agreement with the SM prediction,  placing strong constraints on a variety of new physics scenarios. 

The works~\cite{Blundell1990,Blundell1992,Porsev2009,Porsev2010} used a sum-over-states approach to calculate the PNC $6S_{1/2}-7S_{1/2}$ transition amplitude in \Cs
\begin{equation}\label{SoS_Eqn}
\begin{aligned}
E_{\rm PV} &= \sum_n\left[\frac{\bra{6S_{1/2}}H_W\ket{nP_{1/2}}\bra{nP_{1/2}}{D}_z\ket{7S_{1/2}}}{E_{6S_{1/2}}-E_{nP_{1/2}}}\right.\\
&\left.+\frac{\bra{6S_{1/2}}{D}_{z} \ket{nP_{1/2}} \bra{nP_{1/2}} H_W\ket{7S_{1/2}}}{E_{7S_{1/2}}-E_{nP_{1/2}}}\right]\,.
\end{aligned}
\end{equation}
In this second-order expression, $\ket{nL_J}$ stands for various states of the \Cs\ atom, with 
$n$ being the principal quantum number, $L$ the orbital angular momentum, and $J$ the total angular momentum. These are the true many-body eigen-states of the parity-proper (PP) atomic Hamiltonian. Further, $D_z$ is the $z$-component of the electric dipole operator ${\bf D}\equiv \sum_i \mathbf{d}_i= -\sum_ie{\bf r}_i$  and $H_W = \sum_i h_W(i) $ is the $P$-odd electron-nucleus weak interaction with the single-electron operator $h_W$ having the form
\begin{equation}\label{h_PNC_Def}
	h_W(i)\equiv-\frac{G_F}{2\sqrt{2}}Q_W\gamma_5\rho(r_i)\,.
\end{equation}
Here, $G_F=2.2225\times10^{-14}$ a.u. is the Fermi constant of the weak interaction, $Q_W$ is the weak nuclear charge, $\rho(r)$ is the nuclear neutron density (see Ref.~\cite{derevianko_2001_Breit_NeutronSkin} for a discussion of neutron skin effects) and $\gamma_5$ is the conventional Dirac matrix.

The largest contributions to $E_{\rm PV}$ in the sum-over-states expression~(\ref{SoS_Eqn}) come from terms with $n=6,7,8,9$ (the ``main'' contribution). In Refs.~\cite{Porsev2009,Porsev2010}, the required many-body states were computed using the CC approximation including singles, doubles and valence triples (CCSDvT). The computed CCSDvT wave functions were subsequently used to compute the dipole and weak interaction matrix elements entering Eq.~(\ref{SoS_Eqn}).  Residual contributions to Eq.~(\ref{SoS_Eqn}) come from intermediate states with $n\geq10$ (the ``tail'' contribution) and core-excited states. These residual contributions are sub-dominant and were evaluated using less accurate methods, having an estimated uncertainty of $10\%$.   

In a later work \cite{Dzuba2012}, the value of the residual contributions  was reevaluated and Ref.~\cite{Dzuba2012} claimed a contribution of core-excited states to $E_{\rm PV}$ having an opposite sign
as compared to the analyses of both Refs.~\cite{Blundell1990,Blundell1992} and Refs.~\cite{Porsev2009,Porsev2010}. The \Cs\ weak charge extracted from the revised atomic-structure factor is $1.5\sigma$ away from the SM value, thus relaxing Refs.~\cite{Porsev2009,Porsev2010} constraints on new physics. In addition, Ref.~\cite{Dzuba2012} raised the theoretical uncertainty in the atomic structure factor $k_{\rm PV}$ back to 0.5\%, above the experimental error bar on $E_{\rm PV}$. 

The latest Dalgarno-Lewis-type coupled-cluster computations~\cite{SahDasSoi2021-CsPNC} support both the sign and the value of the core-excited state contributions of Refs.~\cite{Porsev2009,Porsev2010}. However, as of now, a clear understanding of why the two approaches, Refs.~\cite{Dzuba2012} and \cite{Porsev2009,Porsev2010}, lead to core-excited state contribution of opposite signs is still lacking. Furthermore, objections \cite{roberts2021comment} have been raised with regards to the error estimates of 
Ref.~\cite{SahDasSoi2021-CsPNC}.  

It may be observed that the disagreement between Refs.~\cite{Porsev2009,Porsev2010} and \cite{Dzuba2012} arose due to the artificial separation into the ``main'' and ``tail'' contributions characteristic of the sum-over-states method~\cite{WiemanDerevianko-SM50-2019,Safronova2018}. In this paper, we seek to directly include the weak interaction into the single-particle atomic Hamiltonian, thus avoiding this artificial separation, and treat all the intermediate states on equal high-precision footing. In this approach, the single-electron eigen-states of the modified  Hamiltonian will already have a parity-mixed (PM) character. The MBPT calculations of the PM many-body wave functions $\ket{6S_{1/2}'}$ and $\ket{7S_{1/2}'}$ can be carried out in a conventional fashion using this PM singe-electron basis. This PM approach was first suggested in Ref.~\cite{Sandars77} and carried through to all second-order MBPT corrections in Ref.~\cite{Dzuba87}. In this paper, we extend it to a more-complete CC method. Once the PM many-body states are computed, the  PNC amplitude can be expressed simply as,
\begin{equation}
E_{\rm PV} = \bra{6S_{1/2}'}D_z\ket{7S_{1/2}'} \,, \label{Eq:EPV:MP}
\end{equation}
avoiding the summation over intermediate states altogether. 

In addition, the lowest-order Dirac-Hartree-Fock (DHF) result in this PM approach is only 3\% away from the more accurate 
CCSDvT value. This is to be compared with the traditional parity-proper (PP) DHF result which is off by 18\%. This indicates that the correlation corrections in the PM approach are substantially smaller than in the conventional PP method. Depending on the MBPT convergence pattern, one can generically anticipate an improved theoretical accuracy.

Another important point is that in the sum-over-states approach employed in Refs.~\cite{Porsev2009,Porsev2010}, the theoretical uncertainty budget of $E_\mr{PV}$ included comparable contributions of the accurately computed low-lying states (in the CCSDvT approach) and of the
less-accurate highly-excited and core-excited states. Our method would allow us to treat all of these 
contributions on the same high-accuracy CCSDvT footing, thus improving the overall theoretical uncertainty even without the potentially reduced role of the correlation corrections. 

To follow this program through in the context of the CC method, one requires a numerically complete set of PM orbitals (single-particle states) $\{\psi'_i\}$. Generating such PM basis sets and quantifying their numerical accuracy is one of the goals of this paper. A PM basis set has to be obtained in the modified DHF potential of the Xe-like core which includes the  weak interaction~(\ref{h_PNC_Def}). Considering the increased numerical accuracy demanded of the quality of basis sets, we employ the dual-kinetic-balance B-splines basis sets~\cite{Shabaev2004,BELOY2008} which are more numerically robust and have the correct behaviour inside the finite nucleus compared to the B-spline basis sets originally used by the Notre Dame group~\cite{ND_Bset}. 

Once a PM basis set is obtained, one may proceed to computing matrix elements of various operators such as the one-body dipole operator $z'_{ij}$ and the two-body inter-electron Coulomb interaction $g'_{ijkl}$ in the new PM-DHF basis. With these computed matrix elements, the MBPT and CCSDvT expressions can be evaluated. As we will show, all the matrix elements in the PM basis can be decomposed into real and imaginary parts with opposite parities (with the conventional choice of radial wave functions being real-valued). Then all the information about opposite-parity admixtures is contained in the 
imaginary parts of various MBPT expressions. This greatly simplifies the formalism and only requires only minor modifications to already developed and tested MBPT codes. We demonstrate the utility of this technique for the random-phase approximation (RPA) subset of MBPT diagrams and discuss a  strategy for applying these ideas in the more-complete CC calculations.

The paper is organized as follows. In Sec.~\ref{Theory}, we briefly present the set-up of our problem. Although this section does not contain new results, it serves as a starting point for our main discussion and a mean to define our notations. We will also derive, in Sec.~\ref{Sec:ParOpSecondQuantization}, the second-quantized form of the parity operator which will be useful in deriving selection rules for our PM-CC method. Section~\ref{Sec:ParOpSecondQuantization} is followed by Sec.~\ref{PM basis functions}, where we present several methods through which a basis of PM single-electron orbitals may be obtained. In Sec.~\ref{Sec:MatElsPMBasis}, we present the PM matrix elements of one- and two-body operators computed using the obtained PM single-electron orbitals. In Sec.~\ref{Many-body calculations of PNC amplitude}, we illustrate how these PM matrix elements can be used in an RPA calculation of the PNC amplitude. The generalization to the CC method is discussed in Sec.~\ref{Sec:PMCC}. Finally, Sec.~\ref{Conclusions} draws conclusions and presents an outlook for our future work. The paper contains several appendixes which provide further technical details.
Unless specified otherwise, the atomic units, $|e|=m_e=\hbar=1$, are used.

\section{Theory}\label{Theory}
In this section, we lay out the theoretical framework for computing the PNC amplitude in an atom. The material presented here is not new but serves as a starting point and a mean to define our notations.

Let us begin by considering the Hamiltonian of the atomic electrons propagating in the combined PP Coulomb potential $\sum_i V_{\rm nuc}(r_i)$ and the $P$-odd electron-nucleus interaction $H_W=\sum_i h_W(i)$. Here, $i$ labels all the atomic electrons. The full electronic Hamiltonian $H'$ may be decomposed into
\begin{equation}\label{Hamiltonian'}
    \begin{aligned}
        H' &= \sum_i h'_0(i) + V'_c\,,\\
        h'_0(i) &= c\boldsymbol{\alpha}_i\cdot{\bf p}_i+m_ec^2\beta_i\\
        &+V_{\rm nuc}(r_i)+h_W(r_i)+U'(r_i)\,,\\
        V'_c &=\frac12\sum_{i\neq j}\frac{e^2}{|{\bf r}_i-{\bf r}_j|} - \sum_i U'(r_i)\,,
    \end{aligned}
\end{equation}
where $U'(r_i)$ is some single-electron potential to be specified later. We use the prime on $h'_0$ and $V'_c$ to distinguish them from the PP Hamiltonian $h_0=c\boldsymbol{\alpha}_i\cdot{\bf p}_i+m_ec^2\beta_i+V_{\rm nuc}(r_i)+U(r_i)$ and the PP $e-e$ interaction $V_c =\sum_{i\neq j}e^2/(2|{\bf r}_i-{\bf r}_j|) - \sum_i U(r_i)$. For brevity, we suppressed the positive-energy projection operators for the two-electron interactions (no-pair approximation).

As usual, we assume that the energies $\varepsilon'_i$ and orbitals $\psi'_i$ of the unperturbed single-electron Hamiltonian $h'_0$ are known. Note that since the weak interaction is a pseudo-scalar, the total angular momentum $j_i$ and its projection $m_i$ remain good quantum numbers, while the parity is no longer conserved.
For example, $p_{1/2}$ and $s_{1/2}$ orbitals or $d_{5/2}$ and $f_{5/2}$ orbitals of $h_0$ are mixed to form eigen-states of  $h'_0$. 
The many-body eigen-states $\Psi'$ of $H'$ are then expanded over antisymmetrized products of the PM one-particle orbitals $\psi'_i$. In MBPT, one obtains these eigen-states  by treating the residual $e-e$ interaction $V'_c$ as a perturbation.

As the next step, we express the terms in Eq.~\eqref{Hamiltonian'} in second quantization. Let us denote by $a'^\dagger_i$ and $a'_i$ the creation and annihilation operators associated with the one-particle eigen-state $\psi'_i$ of $h'_0$.  We  will follow the indexing convention that core electron orbitals are denoted by the letters at the beginning of alphabet $a,b,c,\dots$, while valence electron orbitals are denoted by $v,w,\dots$, and the indices $i,j,k,\dots$ refer to an arbitrary orbital, core or excited (including valence states). The letters $m,n,p,\dots$ are reserved for those orbitals unoccupied in the core (these could be valence orbitals). 

The operators $H'_0\equiv\sum_ih'_0(i)$ and $V'_c$ may then be written as
\begin{equation}\label{2quant}
    \begin{aligned}
        H'_0 &= \sum_i\varepsilon'_iN[a'^\dagger_ia'_i]\\
        V'_c &=\sum_{ij}\left(V'_{\rm HF}-U'\right)_{ij}N[a'^\dagger_ia'_j]\\
        &+\frac12\sum_{ijkl}g'_{ijkl}N[a'^\dagger_ia'^\dagger_ja'_ka'_l]\,.
    \end{aligned}
\end{equation}
where $N$ denotes normal ordering and $V'_{\rm HF}$ is the PM-DHF potential, whose matrix elements are defined by
\begin{equation}\label{V'HF_def}
	\left(V'_{\rm HF}\right)_{ij}\equiv \sum_{a}\tilde{g}'_{iaja}\,,
\end{equation}
with $\tilde{g}'_{ijkl} \equiv  g'_{ijkl} -  g'_{ijlk}$ being the  anti-symmetrized combination of the Coulomb matrix elements,
\begin{equation}
	g'_{ijkl}\equiv \int\frac{d^3{r}_1 d^3{r}_2}{\left|{\bf r}_1-{\bf r}_2\right|}\psi'^\dagger_i({\bf r}_1)\psi'^\dagger_j({\bf r}_2)\psi'_k({\bf r}_1)\psi'_l({\bf r}_2)\,.
\end{equation}
The irrelevant constant offset energy term $\sum_a\left(V'_{\rm HF}/2-U'\right)_{aa}$ has been omitted in Eq.~\eqref{2quant}.

Notice that the choice $U'=V'_{\rm HF}$ causes the first term in $V'_c$ in Eq.~\eqref{2quant} to vanish, significantly reducing the number of MBPT contributions. In addition, since our final goal is to implement the CCSDvT scheme that has been originally built on DHF potential, we fix $U'=V'_{\rm HF}$. % than that of general parametric potential models.

With $U'=V'_{\rm HF}$ fixed, we now consider the correlation corrections to the independent-particle wave functions. Consider a univalent atom, e.g, \Cs, with a single valence electron above the closed-shell core. The zeroth-order wave function may be expressed as  $\ket{\Psi'^{(0)}_v}=a'^\dagger_v\ket{0'_c}$, where $\ket{0'_c}$ represents the filled Fermi sea of the atomic core (again, the prime indicates that the single-particle orbitals are of PM character). 

To the first order in the residual interaction $V'_c$, the many-body correction $\delta\Psi'_v$ to $\Psi'^{(0)}_v$ has the form
\begin{align}
        \ket{\delta\Psi'_v} &= \sum_{amn}\frac{g'_{nmva}}{\varepsilon'_{av}-\varepsilon'_{nm}}a'^\dagger_na'^\dagger_ma'_a\ket{0'_c} \label{Eq:Psi1}\nonumber\\
        &+\frac12\sum_{abmn}\frac{g'_{nmab}}{\varepsilon'_{ab}-\varepsilon'_{nm}}a'_ba'_aa'^\dagger_na'^\dagger_m a'^\dagger_v \ket{0'_c}\,,
\end{align}
where we used the notation $\varepsilon'_{ij}\equiv\varepsilon'_i+\varepsilon'_j$.
Here, the first term describes a valence electron being promoted to an excited state orbital $m$ with a simultaneous particle-hole excitation of the core (this is so-called valence double excitation, $D_v$, in the language of the CC method). The second contribution is a double particle-hole excitation from the core with valence electron being a spectator (core double excitation, $D_c$). The anti-commutation relations for creation and annihilation operators assure that the  electrons in the second term do not get excited into the valence orbital, i.e. the Pauli exclusion principle is built into the formalism automatically.

With  $\ket{\delta\Psi'_v}$, one can compute the second-order correction to the matrix element of a one-electron operator $T=\sum_{ij}t'_{ij}a'^\dagger_ia'_j$. Once again, the primed quantities refer to the PM orbitals used in computing the matrix elements $t'_{ij} \equiv \langle i'| t | j' \rangle $. The operator $T$ can be, for example, the electric dipole operator $\bf D$. The correction to the matrix element between two valence many-body states $\ket{\Psi'_w}$ and $\ket{\Psi'_v}$, $w\neq v$,  has the form
\begin{align}\label{dtwv}
        \delta T_{wv} &= \bra{\delta\Psi'_w}T\ket{\Psi'_v} + \bra{\Psi'_w}T\ket{\delta\Psi'_v}\nonumber\\
        &=\sum_{an}\frac{t'_{an}\tilde{g}'_{wnva}}{\varepsilon'_a-\varepsilon'_n-\omega}+\sum_{an}\frac{\tilde{g}'_{wavn}t'_{na}}{\varepsilon'_a-\varepsilon'_n+\omega}\,, 
    \end{align}
where $\omega\equiv\varepsilon'_w-\varepsilon'_v$ and $\tilde{g}'_{ijkl}\equiv g'_{ijkl}-g'_{ijlk}$.
Equations~\eqref{Eq:Psi1} and \eqref{dtwv} are, of course, well-known results~\cite{BLUNDELL1987} with the only difference of  using the mixed-parity basis instead of the conventional PP basis.

In Eq.~\eqref{dtwv}, we sum over the core orbitals $a$ and the excited orbitals $n$. Each orbital $\psi'_i$ is characterized by a principal quantum number $n_i$, a total angular momentum $j_i$, and its projection $m_i$. The sums over the magnetic quantum numbers $m_i$ can be carried out analytically using the rules of Racah algebra. Although the sums over $j_i$ are infinite, they are restricted by angular momentum selection rules which radically reduce the number of surviving terms. Moreover, the sums over total angular momenta converge well and in practice, it suffices to sum over a few lowest values of $j_i$. The sums over the principal quantum numbers $n_i$ involve, on the other hand, summing over the infinite discrete spectrum and integrating over the continuum. In the basis set method, these infinite summations are replaced by summations over a finite-size pseudo-spectrum \cite{Johnson1988,Bachau2001,Shabaev2004,BELOY2008}.

The basis orbitals in the pseudo-spectrum are obtained by placing the atom in a sufficiently large cavity and imposing boundary conditions at the cavity wall and at the origin (see Ref.~\cite{BELOY2008} for further details on a dual-kinetic-basis B-spline sets used in our paper). For each value of $j_i$, one then finds a discrete set of $2N$ orbitals, $N$ from the Dirac sea and the remaining $N$ with energies above the Dirac sea threshold (conventionally referred to as ``negative'' and ``positive'' energy parts of the spectrum in analogy with free-fermion solutions). 

If the size of the cavity is large enough, typically about $40a_0/Z$ where $a_0$ is the Bohr radius, the low-lying orbitals with positive energies map with a good accuracy to the discrete orbitals of the exact DHF spectrum. Higher-energy orbitals do not closely match their physical counterparts. Nevertheless, since the pseudo-spectrum is complete, it forms a basis set for the function space spanning the cavity and thus can be used instead of the real spectrum to evaluate correlation corrections to states confined to the cavity. From now on, all single-particle orbitals $\psi'_i$ are understood to be members of the B-spline basis set.

To reiterate, the parity-mixed (PM) formalism presented so far is essentially the same as in the conventional parity-proper (PP) MBPT. The only difference is that all the quantities are defined with respect to the PM orbitals $\psi'_i$ instead of the PP ones. Since PM orbitals are eigen-states of total angular momentum, one can directly use the results of angular reduction for various MBPT expressions, and the existing MBPT codes require minor changes, mostly related to modifying parity selection rules and the use of Coulomb integrals in the PM basis. In Sec.~\ref{Sec:ParOpSecondQuantization}, we derive the second-quantized form of the parity operator which will be useful for deriving the PM selection rules and in Sec.~\ref{PM basis functions}, we present several methods through which the PM orbitals may be generated in practice.

\section{Parity operator in second quantization}
\label{Sec:ParOpSecondQuantization}
Since the MBPT derivations are built on the second-quantization formalism, in this section  we  derive the second-quantized form of the parity operator $\Pi$ to be used in deriving parity selection rules (see Appendix \ref{App_B}). Parity transformation is 
defined by $\mathbf{r}_i \rightarrow -\mathbf{r}_i$ for all the $N_e$ electrons in the system. Consider a PP state (Slater determinant) $\ket{\Psi^{m_1\ldots m_\mu}_{a_1\ldots a_\mu}}$ composed of orbitals of definite parity. This many-body state is obtained by removing $\mu=0,\ldots,N_e$ electrons $a_1,\ldots,a_\mu$ from the reference state $\ket{\Psi^{(0)}_v}$ while adding the same number of excited electrons $m_1,\ldots,m_\mu$. Notice that in this notation the valence orbital is treated as initially occupied and, thereby, $v$ can be one of the labels  $a_1,\ldots,a_\mu$. In the second quantization,
\begin{equation}\label{Eq:Psi^mm_aa}
    \ket{\Psi^{m_1\ldots m_\mu}_{a_1\ldots a_\mu}}\equiv a^\dagger_{m_1}\ldots a^\dagger_{m_\mu}a_{a_1}\ldots a_{a_\mu}\ket{\Psi^{(0)}_v}\,.
\end{equation}
We emphasize that the creation and annihilation operators in Eq.~\eqref{Eq:Psi^mm_aa} are the PP ones.

Since the PP Hamiltonian $H_0\equiv\sum_i h_0(i)$ is invariant under spatial reflection, it commutes with the parity operator,$[H_0,\Pi]=0$. As a result, the states $\ket{\Psi_v^{(0)}}$ and $\ket{\Psi^{m_1\ldots m_\mu}_{a_1\ldots a_\mu}}$, being eigen-states of $H_0$, are also an eigen-states of the parity operator $\Pi$. Furthermore, since $\ket{\Psi_v^{(0)}}$ and $\ket{\Psi^{m_1\ldots m_\mu}_{a_1\ldots a_\mu}}$ are antisymmetrized products of single-electron orbitals, their eigen-values with respect to $\Pi$ equal the products of the parities of their constituents
\begin{subequations}
	\begin{align}
		\Pi \ket{\Psi_v^{(0)}} &= (-1)^{\ell_v}\ket{\Psi_v^{(0)}}\,,\label{Eq:PiEigenValue00}\\
		\Pi \ket{\Psi^{m_1\ldots m_\mu}_{a_1\ldots a_\mu}} &= (-1)^{\ell_v+\sum_{i=1}^\mu\ell_{a_i}+\ell_{m_i}} \ket{\Psi^{m_1\ldots m_\mu}_{a_1\ldots a_\mu}}\,, \label{Eq:PiEigenValue}
	\end{align}
\end{subequations}
where we have used the fact that the closed-shell core has even parity. 

To transform an operator into the second-quantized form, we recall the conventional formula
\begin{equation}
    \Pi = \sum_{\alpha,\beta} \ket{\alpha} \langle \alpha |\Pi | \beta \rangle \bra{\beta} \,.
\end{equation}
Its proof  relies on the  identity resolution (closure relation) for a complete  orthonormal basis $I=\sum_\alpha \ket{\alpha} \bra{\alpha} = \sum_{\beta} \ket{\beta} \bra{\beta}$. For a system of identical particles, however, one needs to proceed with caution due to the possibility of  permutations of orbitals in   $\ket{\Psi^{m_1\ldots m_\mu}_{a_1\ldots a_\mu}}$. Indeed, in the many-fermion case, the orthonormality condition reads
\begin{align}
\braket{\Psi^{m_1\ldots m_\mu}_{a_1\ldots a_\mu}|\Psi^{n_1\ldots n_\nu}_{b_1\ldots b_\nu}}= \delta_{\mu\nu}  \delta^{b_1\ldots b_\nu}_{a_1\ldots a_\mu}\delta^{n_1\ldots n_\nu}_{m_1\ldots m_\mu}\,, \label{Eq:Ortho}
\end{align}
where the generalized Kronecker delta is defined as
\begin{equation}\label{Pi_mel}
    \delta^{k_1\ldots k_\mu}_{l_1\ldots l_\mu} = \left\{\begin{matrix}
        +1 & \begin{matrix}
            k_1,\ldots, k_\mu {\rm\,\,are\,\,an\,\,even}\\
            {\rm \quad permutation\,\,of\,\,} l_1,\ldots, l_\mu
        \end{matrix}\\
        -1 & \begin{matrix}
            k_1,\ldots, k_\mu {\rm\,\,are\,\,an\,\,odd}\\
            {\rm \quad permutation\,\,of\,\,} l_1,\ldots, l_\mu
        \end{matrix}\\
        0 & {\rm otherwise}
    \end{matrix}\right.\,.
\end{equation}

For many-fermion systems, the general closure relation is given in Ref.~\cite{Arponen1987}. Since $\Pi$ is a diagonal operator, the general identity resolution~\cite{Arponen1987} simplifies to
\begin{align}\label{Closure}
    I=\sum_{\mu=0}^{N_e}\frac{1}{(\mu!)^2}\sum_{\{a\}\{m\}}\ket{\Psi^{m_1\ldots m_\mu}_{a_1\ldots a_\mu}}\bra{\Psi^{m_1\ldots m_\mu}_{a_1\ldots a_\mu}}\,,
    % \begin{smallmatrix}
    % a_1,\ldots,a_i\\
    % m_1,\ldots,m_j
    % \end{smallmatrix}
\end{align}
where $\{a\}$ and $\{m\}$ denote strings of orbital labels in $\ket{\Psi^{m_1\ldots m_\mu}_{a_1\ldots a_\mu}}$.

% In order to put any operator in second-quantized form, we first need an expression for the identity operator $I$, i.e., the closure relation. Since we are concerned with the CC method, the most convenient form of $\Pi$ will contain particle-hole creation and annihilation operators with respect to the reference state $\Psi_v^{(0)}$ consisting of an electron above the filled Fermi sea of the atomic core. Furthermore, since the wave operator $\Omega$ conserves the total number of electrons $N_e$, we need to care only about the resolution of the identity operator restricted to the subspace spanned by states with $N_e$ occupied orbitals. 

% Within the particle-hole framework, the closure relation may be thus written as \cite{Arponen1987}
% \begin{align}\label{Closure}
%     I=\sum_{\mu=0}^{N_e}\frac{1}{(\mu!)^2}\sum_{\{a\}\{m\}}\ket{\Psi^{m_1\ldots m_\mu}_{a_1\ldots a_\mu}}\bra{\Psi^{m_1\ldots m_\mu}_{a_1\ldots a_\mu}}\,,
%     % \begin{smallmatrix}
%     % a_1,\ldots,a_i\\
%     % m_1,\ldots,m_j
%     % \end{smallmatrix}
% \end{align}
% where
% \begin{equation}
%     \ket{\Psi^{m_1\ldots m_\mu}_{a_1\ldots a_\mu}}\equiv a^\dagger_{m_1}\ldots a^\dagger_{m_\mu}a_{a_1}\ldots a_{a_\mu}\ket{\Psi^{(0)}_v}\,,
% \end{equation}
% is the state obtained by removing $\mu=0,\ldots,N_e$ electrons $a_1,\ldots,a_\mu$ from the reference state $\Psi^{(0)}_v$ while adding the same number of excited electrons $m_1,\ldots,m_\mu$. The expressions $\{a\}$ and $\{m\}$ are short for the (un-ordered) sets $\{a_1,\ldots,a_\mu\}$ and $\{m_1,\ldots,m_\mu\}$, respectively.

Sandwiching $\Pi$ in between two identity operators and using the closure relation~\eqref{Closure}, we find
\begin{align}\label{Pi_2nd_quant}
   \Pi &= \sum_{\mu,\nu=0}^{N_e}\frac{1}{(\mu!\nu!)^2}\sum_{\{a\}\{m\}}\sum_{\{b\}\{n\}}\nonumber\\
    &\times\bra{\Psi^{m_1\ldots m_\mu}_{a_1\ldots a_\mu}}\Pi\ket{\Psi^{n_1\ldots n_\nu}_{b_1\ldots b_\nu}}\ket{\Psi^{m_1\ldots m_\mu}_{a_1\ldots a_\mu}}\bra{\Psi^{n_1\ldots n_\nu}_{b_1\ldots b_\nu}}\,.
\end{align}
Using the eigenvalue equation \eqref{Eq:PiEigenValue}, we obtain the following result for the matrix element of $\Pi$
\begin{align}
    \bra{\Psi^{m_1\ldots m_\mu}_{a_1\ldots a_\mu}}\Pi\ket{\Psi^{n_1\ldots n_\nu}_{b_1\ldots b_\nu}} &= (-1)^{\ell_v+\sum_{i=1}^\nu\ell_{b_i}+\ell_{n_i}}\nonumber\\
    &\times\delta_{\mu \nu} \delta^{b_1\ldots b_\nu}_{a_1\ldots a_\mu}\delta^{n_1\ldots n_\nu}_{m_1\ldots m_\mu}\,,
\end{align}
where we used the orthonormality relation~\eqref{Eq:Ortho}.
Substituting Eq.~\eqref{Pi_mel} into Eq.~\eqref{Pi_2nd_quant}, we finally obtain
\begin{align}\label{Eq:18}
   \Pi&= \sum_{\mu=0}^{N_e}\frac{1}{(\mu!)^2}\sum_{\{a\}\{m\}}(-1)^{\ell_v+\sum_{i=1}^\mu\ell_{a_i}+\ell_{m_i}}\nonumber\\
    &\times a^\dagger_{m_1}\ldots a^\dagger_{m_\mu}a_{a_1}\ldots a_{a_\mu}\ket{\Psi^{(0)}_v}\bra{\Psi^{(0)}_v}\nonumber\\
    &\times a^\dagger_{a_\mu}\ldots a^\dagger_{a_1}a_{m_\mu}\ldots a_{m_1}\,.
\end{align}

\section{Parity-mixed single-electron basis orbitals}
\label{PM basis functions}
In this section, we demonstrate how the PM single-particle wave functions $\psi'_i\equiv\psi'_{n_ij_im_i}$ may be obtained. They are the solutions to the PM-DHF equation
\begin{equation}\label{DHF-PNC-SPL-EQN}
\begin{aligned}
&h_0' \psi'_{n_ij_im_i}=\varepsilon'_{n_ij_i}\psi'_{n_ij_im_i}\,, \\
    &h_0'= c\boldsymbol{\alpha} \cdot{\bf p} +m_ec^2\beta + 
    V_{\rm nuc}+h_W+V'_\mathrm{HF} \,.
\end{aligned}
\end{equation}
Here, $n_i$ is the principal quantum number, $j_i$ is the total angular momentum and $m_i$ is the projection 
of $j_i$ on a quantization axis.

Our goal is to expand $\psi'_i$ in terms of the PP orbitals $\psi^P_i\equiv\psi^P_{n_i\ell_ij_im_i}$, which are solutions to the conventional DHF equation
\begin{equation}
    \begin{aligned}\label{Normal_DHF}
    &h_0\psi^P_{n_i\ell_ij_im_i}=\varepsilon_{n_i\ell_ij_i}\psi^P_{n_i\ell_ij_im_i}\,, \\
       &h_0= c\boldsymbol{\alpha} \cdot{\bf p} +m_ec^2\beta + 
    V_{\rm nuc}+ V_\mathrm{HF} \,.
\end{aligned}
\end{equation}
Note that besides the principal quantum number $n_i$, the total angular momentum $j_i$, and the magnetic quantum number $m_i$, we have characterized the PP orbital $\psi^P_i$ with an extra quantum number, the orbital angular momentum $\ell_i$, which indicates that $\psi^P_i$ has a definite parity $(-1)^{\ell_i}$.

The two DHF potentials $V'_\mathrm{HF}$ and $V_\mathrm{HF}$  depend on the core orbitals.
Since core orbitals are self-consistent solutions of these DHF equations, the PM and PP core orbitals differ. Therefore, the effects of the weak interaction on the single-electron orbitals are contained in the difference of the two Hamiltonians, 
\begin{equation}
    \Delta h = h_0' -h_0 = h_W +V'_\mathrm{HF} -  V_\mathrm{HF} \,,\label{Eq:DeltaH}
\end{equation}
which is a pseudo-scalar interaction, preserving rotational symmetry but spoiling mirror symmetry (this is why we have used the quantum numbers $n_i$, $j_i$ and $m_i$ but not $\ell_i$ to index the PM orbital $\psi'_i$).  
This suggests the following parameterization~\cite{BluJohSap1998.ChapterInBook} for the solutions to the PM-DHF equations \eqref{DHF-PNC-SPL-EQN},
\begin{subequations}\label{PNC_DHF_para}
    \begin{align}
    \psi'_i(\mathbf{r}) &=   \psi_i(\mathbf{r})+\im\eta\bar{\psi}_i(\mathbf{r}) \,, \label{opp_par_comb}\\
    \psi_i(\mathbf{r}) & =\frac{1}{r}\left(\begin{array}{c}\im P_{n_i\kappa_i}(r) \Omega_{\kappa_i m_i}(\hat{\mathbf{r}}) \\ Q_{n_i\kappa_i}(r) \Omega_{-\kappa_i m_i}(\hat{\mathbf{r}})\end{array}\right) \,,\label{parameterize}\\
    \bar{\psi}_i(\mathbf{r}) &=  \frac{1}{r}\left(\begin{array}{c}\im \bar{P}_{n_i\kappa_i}(r) \Omega_{-\kappa_i m_i}(\hat{\mathbf{r}}) \\ \bar{Q}_{n_i\kappa_i}(r) \Omega_{\kappa_i m_i}(\hat{\mathbf{r}})\end{array}\right)\label{psi_bar}\,,
    \end{align}
\end{subequations}
where
\begin{equation}
	\eta\equiv \frac{G_FQ_W}{2\sqrt{2}a_0^2}\sim10^{-15}
\end{equation} 
is a dimensionless factor characteristic of the strength of the weak interaction and its numerical value is given for \Cs. In Eqs.~\eqref{PNC_DHF_para}, $\kappa_i=\left(j_i+\frac{1}{2}\right)(-1)^{j_i+\ell_i+1/2}$ is a relativistic angular quantum number that encodes the values of both the total and orbital angular momenta $j_i$ and $\ell_i$. From the definition of the relativistic angular quantum number, flipping the parity $(-1)^{\ell_i}$ of the orbital while preserving the total angular momentum $j_i$ is equivalent to changing $\kappa_i \rightarrow -\kappa_i$, as presented in the parameterization of $\bar{\psi}_i$, Eq.~\eqref{psi_bar}.

It is appropriate to pause here and introduce a point of semantics. Although the orbital $\psi'_i$ does not have a definite parity, one can nevertheless speak of its ``nominal parity'', defined as that of the component $\psi_i$, which is not suppressed by the factor $\eta$. In the light of Eq.~\eqref{opp_par_comb}, we shall refer to $\psi_i$ as the ``real'' component and $\bar{\psi}_i$ as the ``imaginary'' component of $\psi'_i$. In what follows, in particular when discussing MBPT and the CC formalism, we shall refer to the nominal parity of a PM single-electron orbital, meaning that of its real component. The nominal parity of $\psi'_i$ is thus $(-1)^{\ell_i}$.

The combination \eqref{opp_par_comb} clearly demonstrates the admixing of the opposite-parity orbital $\bar{\psi}_i$ with $\kappa_i \rightarrow -\kappa_i$ to the reference orbital $\psi_i$. With the imaginary unity factored out in the admixture component
$\bar{\psi}_i$ and with the conventional definition of the spherical spinors $\Omega_{\kappa m}$~\cite{WRJBook}, all the radial wave functions $P_{n_i\kappa_i}$, $Q_{n_i\kappa_i}$, $\bar{P}_{n_i\kappa_i}$, and $\bar{Q}_{n_i\kappa_i}$ can be chosen to be real-valued. The rest of this section will be devoted to finding these radial components. In what follows, we will assume a parameteriziton for the PP solutions to Eq.~\eqref{Normal_DHF} similar to that presented in Eqs.~\eqref{parameterize}, namely
\begin{equation}\label{parameterize_norm}
    \psi^P_i(\mathbf{r}) =\frac{1}{r}\left(\begin{array}{c}\im P^P_{n_i\kappa_i}(r) \Omega_{\kappa_i m_i}(\hat{\mathbf{r}}) \\ Q^P_{n_i\kappa_i}(r) \Omega_{-\kappa_i m_i}(\hat{\mathbf{r}})\end{array}\right) \,.
\end{equation}

Where there is no risk of confusion, we will abbreviate $P_{n_i\kappa_i}$ ($Q_{n_i\kappa_i}$) to $P_i$ ($Q_i$) and $P^P_{n_i\kappa_i}$ ($Q^P_{n_i\kappa_i}$) to $P^P_i$ ($Q^P_i$). The energy eigenvalues $\epsilon'_{n_ij_i}$ and $\epsilon_{n_il_ij_i}$ in Eqs.~\eqref{DHF-PNC-SPL-EQN} and \eqref{Normal_DHF} will also be abbreviated to $\epsilon'_i$ and $\epsilon_i$, respectively.

\subsection{Parity-mixed basis set construction: finite-difference method}\label{Dalgarno_Lewis}

We start our computation of the radial functions $P_i$, $Q_i$, $\bar{P}_i$ and $\bar{Q}_i$ with a discussion of the finite-difference method, where we integrate the PM-DHF equations directly,
without using the basis set technique. While the finite-difference method does not produce the finite basis set for MBPT-type calculations, it generates the PM core orbitals entering the PM-DHF potential that can be used in constructing the basis set. In addition, the finite-difference method provides reference results that are used to gauge the fidelity of the basis set representation of core and low-energy orbitals.

Due to the smallness of the dimensionless coupling constant $\eta$, we may set $\psi_i=\psi^P_i$ which is accurate up to $O(\eta^2)$. As a result, to the first order in $\eta$, Eq.~\eqref{DHF-PNC-SPL-EQN} yields a pair of integro-differential equations for the radial functions $\bar{P}_i$ and $\bar{Q}_i$. For a core orbital, $i=a$, these equations read
\begin{subequations}\label{DL_eqns}
    \begin{align}
       &c\left(\frac{d}{dr}-\frac{\kappa_a}{r}\right)\bar{P}_a-\left(V_{\rm eff}-\varepsilon_a-c^2\right)\bar{Q}_a\nonumber\\
       &= -\rho_{\rm nuc}P^P_a - \sum_b \bar{V}_{ba}Q^P_b - \sum_b V_{ba}\bar{Q}_b\,,\\
       &c\left(\frac{d}{dr}+\frac{\kappa_a}{r}\right)\bar{Q}_a+\left(V_{\rm eff}-\varepsilon_a+c^2\right)\bar{P}_a\nonumber\\
       &= -\rho_{\rm nuc}Q^P_a + \sum_b \bar{V}_{ba}P^P_b + \sum_b V_{ba}\bar{P}_b\,,
    \end{align} 
\end{subequations}
where $V_{\rm eff}$ is an effective potential comprising of the electron-nucleus Coulomb potential and the direct part of the conventionally-defined DHF potential ($[j] \equiv 2j+1$) 
\begin{equation} \label{Eq:V_dir}
    V_{\rm eff}(r) \equiv V_{\rm nuc}(r) + \sum_b[j_b]v_0(b,b,r)\,,
\end{equation}
while $V_{ba}$ is the  DHF exchange potential 
\begin{equation}\label{Vai_def}
    V_{ba}(r) \equiv \frac{1}{[j_a]}\sum_k\rmel{\kappa_a}{C_k}{\kappa_b}^2v_k(b,a,r)\,, 
\end{equation}
and $\bar{V}_{ba}$ is the PNC-DHF exchange potential
\begin{equation}\label{Vai_bar_def}
    \bar{V}_{ba}(r) \equiv \frac{1}{[j_a]}\sum_k\rmel{-\kappa_a}{C_k}{\kappa_b}^2\left(v_k(b,\bar{a},r)-v_k(\bar{b},a,r)\right)\,.
\end{equation}
In Eqs.~(\ref{Eq:V_dir}) and (\ref{Vai_def}), the multipolar potential $v_k(b,a,r)$ is defined as
\begin{equation}
    \begin{aligned}
        v_k(b,a,r) &\equiv \int r_<^k r_>^{-k-1}dr'\\
        &\times \left(P^P_b(r')P^P_a(r')+Q^P_b(r')Q^P_a(r')\right)\,,
    \end{aligned}
\end{equation}
whereas the quantity $v_k(b,\bar{a},r)$ in Eq.~\eqref{Vai_bar_def} is defined as
\begin{equation}
    \begin{aligned}
        v_k(b,\bar{a},r) &\equiv \int r_<^k r_>^{-k-1}dr'\\
        &\times \left(P^P_b(r')\bar{P}_a(r')+Q^P_b(r')\bar{Q}_a(r')\right)\,,
    \end{aligned}
\end{equation}
and similarly for $v_k(\bar{b},a,r)$. In these equations, $r_< = \min(r,r')$ and $r_> = \max(r,r')$.

Equations~\eqref{DL_eqns} may be solved using an iterative scheme ($n$ is the iteration number)
\begin{subequations}\label{DL_eqns_iter}
	\begin{align}
		c\left(\frac{d}{dr}-\frac{\kappa_a}{r}\right)\bar{P}^{(n+1)}_a&-\left(V_{\rm eff}-\varepsilon_a-c^2\right)\bar{Q}^{(n+1)}_a\nonumber\\
		&= -\rho_{\rm nuc}P^P_a - X_a^{(n+1)}\,,\\
		c\left(\frac{d}{dr}+\frac{\kappa_a}{r}\right)\bar{Q}^{(n+1)}_a&+\left(V_{\rm eff}-\varepsilon_a+c^2\right)\bar{P}^{(n+1)}_a\nonumber\\
		&= -\rho_{\rm nuc}Q^P_a + Y_a^{(n+1)}\,,
	\end{align} 
\end{subequations}
where we have defined
\begin{subequations}\label{XY_def}
	\begin{align}
		X_a^{(n+1)} \equiv \sum_b \bar{V}^{(n+1)}_{ba}Q^P_b - \sum_b V_{ba}\bar{Q}^{(n+1)}_b\,,\\
		Y_a^{(n+1)} \equiv \sum_b \bar{V}^{(n+1)}_{ba}P^P_b + \sum_b V_{ba}\bar{P}^{(n+1)}_b\,.
	\end{align}
\end{subequations}
Note that $X_a$ and $Y_a$ are themselves functions of $\bar{P}_a$ and $\bar{Q}_a$, which appear explicitly in the second terms of Eqs.~\eqref{XY_def} and implicitly via the PNC-DHF exchange potential $\bar{V}_{ba}$ in the first terms of Eqs.~\eqref{XY_def}.

Equations \eqref{DL_eqns_iter} are inhomogeneous second-order differential equations which may be solved using the conventional technique of variation of parameters. In this method, one first finds the solution to the homogeneous version of Eqs.~\eqref{DL_eqns_iter}. Since the operators acting on $\bar{P}_a$ and $\bar{Q}_a$ on the left hand side of Eqs.~\eqref{DL_eqns_iter} do not change from iteration to iteration, neither will the homogeneous solutions. As a result, they only need to be computed once. The inhomogeneous solutions $\bar{P}^{(n+1)}_a$ and $\bar{Q}^{(n+1)}_a$ are then obtained by convoluting the corresponding homogeneous solutions with the right hand sides of Eqs.~\eqref{DL_eqns_iter} (see, e.g., Ref.~\cite{WRJBook} for further details on the technique of variation of parameters for DHF equation).

Once the radial functions $\bar{P}_a$ and $\bar{Q}_a$ are obtained, we may proceed to solving for the radial functions $\bar{P}_m$ and $\bar{Q}_m$ of the unoccupied orbitals. The equations for $\bar{P}_m$ and $\bar{Q}_m$ are obtained by replacing $a\rightarrow m$ in Eqs.~\eqref{DL_eqns} and we may set up a similar iteration scheme for valence orbitals as in Eqs.~\eqref{DL_eqns_iter}. Note that in this case, the driving terms $X_m$ and $Y_m$ depend on $\bar{P}_m$ and $\bar{Q}_m$ via the PNC-DHF potential $\bar{V}_{bm}$ only. Other than this, the procedure for solving the PNC-DHF for unoccupied orbitals is the same as for core orbitals.

We note, however, that in general, the iteration scheme \eqref{DL_eqns_iter} and also its counterpart for unoccupied orbitals do not converge but oscillate. Such behavior can be removed if the driving terms $X$ and $Y$ are changed slowly between iterations. This is accomplished by setting
\begin{equation}
	\begin{aligned}
		X_i^{(n+1)} &= \lambda X_i^{(n+1)} + (1-\lambda) X_i^{(n)}\,,\\
		Y_i^{(n+1)} &= \lambda Y_i^{(n+1)} + (1-\lambda) Y_i^{(n)}\,.
	\end{aligned}
\end{equation}
We find that choosing $\lambda= 0.01 \sim 0.1$ generally assures iteration convergence for all the orbitals, core and unoccupied. In the rest of this section, we shall discuss two matrix methods which allow us to avoid altogether this issue of convergence.

As a check for our numerical procedure for the finite-difference method, we recovered the previous literature results~\cite{Blundell1992,Johnson2001} for the lowest order $6S_{1/2}-7S_{1/2}$ PNC transition amplitude in \Cs. We calculated the amplitudes in both the frozen-core (fc) approximation, which involves neglecting the PNC effects on core orbitals, obtaining
\begin{equation}\label{fin_dif_fro_cor_res}
	E^{\rm fc}_{\rm PV} = 0.73946\times10^{-11}\im|e|a_0(Q_W/N)\,,
\end{equation} 
and the full core-perturbed (cp) case, where the PNC perturbation to core orbitals is fully taken into account, obtaining
\begin{equation}\label{fin_dif_res}
	E^{\rm cp}_{\rm PV} = 0.92700\times10^{-11}\im|e|a_0(Q_W/N)\,.
\end{equation}  
%\APD{ do you need cp superscript in the above?\\}

In all our numerical examples, the nuclear charge distribution is approximated by a Fermi distribution $\rho_\mr{nuc}(r)=\rho_0/(1+\exp[(r-c)/a])$, where $\rho_0$ is a normalization constant. For \Cs, we use $c=5.6748 \, \mr{fm}$ and $a =0.52338 \, \mr{fm}$. We also use the same nuclear distribution in computations of weak interaction~(\ref{h_PNC_Def}), $\rho(r) \equiv \rho_\mr{nuc}(r)$.

\subsection{Parity-mixed basis set construction: exact matrix diagonalization methods}\label{Matrix diagonalization method}
The goal of this section is to construct a PM-DHF basis set $\{\psi'_i\}$  by transforming  
 a numerically complete PP-DHF basis set $\{\psi_i^P\}$: $\{\psi_i^P\} \rightarrow \{\psi'_i\}$ (basis rotation). The PP-DHF basis 
sets based on the solution of the conventional PP-DHF equations are widely used both in atomic structure  and quantum chemistry calculations and we assume that the set $\{\psi_i^P\}$ was pre-computed.

The two DHF equations, PM- and PP-DHF, differ by $\Delta h$, Eq.~\eqref{Eq:DeltaH}, which includes the weak interaction and the difference between the two DHF potentials.
While the weak interaction is a small perturbation, $\Delta h\sim \eta \sim 10^{-15}$ for \Cs, one may encounter accidental degeneracies between basis orbitals of opposite parities (especially in the high-energy part of the pseudo-spectra), making application of perturbative approaches error-prone.
In this subsection, we discuss two exact methods based on the diagonalization of the PM-DHF Hamiltonian,
and in the next subsection, we explore the perturbative approach. 

The two approaches considered in this subsection involve transforming the PP-DHF basis $\{\psi_i^P\}$ into the desired PM-DHF basis $\{\psi'_i\}$: (i) without requiring the prior computation of the PM-DHF core orbitals and (ii) with the PM-DHF potential pre-computed using, say, the finite-difference method of the previous section.

Let us consider the first method. Suppose we do not know the PM-DHF core orbitals and thus can not immediately construct the PM-DHF potential beforehand. Recall that the PM-DHF orbitals are represented as  $\psi'_i = \psi_i +i\eta\bar{\psi}_i$,~Eq.~\eqref{PNC_DHF_para},
where $\psi_i$ is the nominal parity contribution and $\bar{\psi}_i$ is the opposite-parity admixture.  Since the PP-DHF set $\{\psi_i^P\}$ forms a numerically-complete basis, the nominal parity contribution $\psi_i$ can be expanded in terms of the $2N$ orbitals $\psi^P_j$ of the same total angular momentum and parity as $\psi_i$, i.e. $\kappa_j=\kappa_i$ (recall that $2N$ is the number of basis functions for a given $\kappa$). Similarly, the opposite-parity admixtures $\bar{\psi}_i$ may be expanded over the $2N$ PP-DHF orbitals $\psi^P_{\bar j}$ which have the same total angular momentum but opposite parity to $\psi_i$, i.e. $\kappa_{\bar j}=-\kappa_i$. 

As a result, the PM wave function $\psi'_i$ may be written as
\begin{equation}\label{psi'_expansion}
	\psi'_i = \sum_j \chi_{ij}\psi^P_j + \im\sum_{\bar j}\chi_{i\bar j}\psi^P_{\bar j}\,,
\end{equation}
where the factor $\eta$ has been absorbed into the opposite-parity admixture coefficients $\chi_{i\bar{j}}$, i.e. $\chi_{i\bar{j}} \sim O(\eta)$. More explicitly, Eq.~\eqref{psi'_expansion} reads
\begin{equation}\label{expansion_explicit}
    	\psi'_{n_ij_im_i} = \sum_{n_j} \chi_{ij}\psi^P_{n_j\ell_ij_im_i} + \im\sum_{n_{\bar{j}}}\chi_{i\bar j}\psi^P_{n_{\bar{j}}\bar{\ell}_ij_im_i}\,,
\end{equation}
where the index $\bar{\ell}_i=\ell_i\pm1$ indicates that terms in the second summation in Eq.~\eqref{expansion_explicit} have opposite parities to those in the first summation.

In terms of the radial wave functions $P_i$ ($Q_i$) and $\bar{P}_i$ ($\bar{Q}_i$), Eqs.~\eqref{psi'_expansion} and \eqref{expansion_explicit} are equivalent to
\begin{equation}
    \begin{aligned}
         P_i&=\sum_{n_j}\chi_{ij}P^P_{n_j\kappa_i}\,, \quad G_i=\sum_{n_j}\chi_{ij}Q^P_{n_j\kappa_i}\,, \\
         \bar{P}_i&=\sum_{n_j}\chi_{i\bar{j}}P^P_{n_j-\kappa_i}\,, \quad \bar{Q}_i=\sum_{n_j}\chi_{i\bar{j}}Q^P_{n_j-\kappa_i}\,,
    \end{aligned}
\end{equation}
where we have fixed the relativistic angular numbers $\kappa_j = \pm \kappa_i$ to reflect parities.

Substituting the expansion \eqref{psi'_expansion} into Eq.~\eqref{DHF-PNC-SPL-EQN}, multiplying with ${\psi^P_j}^\dagger$ and ${\psi^P_{\bar j}}^\dagger$, and  then integrating, we obtain
\begin{subequations}\label{explicit_Eig_Eqn}
	\begin{align}
		\varepsilon_k\chi_{ij} + \im\sum_{\bar j}\bra{j}\Delta h\ket{\bar j}\chi_{i\bar j} = \varepsilon'_i\chi_{ij}\,,\label{ei1} \\
		\sum_j\bra{\bar j}\Delta h\ket{j}\chi_{ij} + \im\varepsilon_{\bar j}\chi_{i{\bar j}} = \im\varepsilon'_i \chi_{i{\bar j}}\,,\label{ei2}
	\end{align}
\end{subequations}
where we have used the fact that $\Delta h$, Eq.~(\ref{Eq:DeltaH}), can only connect orbitals of opposite parities. 

Equations \eqref{explicit_Eig_Eqn} may be put in the form of an eigenvalue matrix equation,
\begin{equation}\label{eig_eqn}
	\bs{M}\boldsymbol{\chi}_i = \varepsilon'_i\boldsymbol{\chi}_i\,,
\end{equation}
where $\boldsymbol{\chi}_i \equiv (\chi_{ij},\chi_{i{\bar j}})$ and $\bs{M}$ is a $4N\times4N$ matrix defined by
\begin{equation}\label{M_def}
	\begin{aligned}
		M_{jj} &= \varepsilon_j\,,\quad M_{j\bar{j}} = \im\bra{j}\Delta h\ket{\bar j}\,,\\
		M_{{\bar j}{\bar j}} &= \varepsilon_{\bar j}\,,\quad M_{\bar{j}j} = -\im\bra{\bar j}\Delta h\ket{j}\,.
	\end{aligned}
\end{equation}

The matrix $\bs{M}$, Eq.~\eqref{M_def}, is a real symmetric matrix. As a result, its eigenvalues $\varepsilon'_i$ and eigenvectors $\boldsymbol{\chi}_i$ are real. Further, we may express the off-diagonal elements of  $\bs{M}$ in a more explicit form:
\begin{align}\label{M_off_diag}
		-\im\bra{\bar{j}}\Delta h\ket{j} &=-\im \bra{\bar{j}}h_W+V'_{\rm HF}-V_{\rm HF}\ket{j}\nonumber\\
		&= \eta S_{\bar{j}j}-\sum_{ak\bar{k}}\chi_{ak}\chi_{a\bar{k}}(g_{\bar{j}k\bar{k}j}-g_{\bar{j}\bar{k}kj})\,,
\end{align}
where
\begin{equation}\label{S_def}
    S_{ij} \equiv\bra{i}\im\rho\gamma_5\ket{j}= \int\left(P^P_iQ^P_j-P^P_jQ^P_i\right)\rho(r) dr\,.
\end{equation}

Note that here, the Coulomb matrix elements $g_{\bar{j}k\bar{k}j}$ and $g_{\bar{j}\bar{k}kj}$ are defined with respect to the PP basis orbitals $\psi^P_j$, $\psi^P_k$, $\psi^P_{\bar{k}}$ and $\psi^P_{\bar{j}}$. The orbital $\psi^P_k$ has the same total angular momentum and parity as the core orbital $\psi^P_a$, i.e. $\kappa_k=\kappa_a$, whereas $\psi^P_{\bar{k}}$ has the same total angular momentum but opposite parity to $\psi^P_a$, i.e. $\kappa_{\bar k}=-\kappa_a$. Note that the orbitals $\psi^P_k$ and $\psi^P_{\bar k}$ are not limited to the core and do not necessarily have the same principal quantum numbers. The quantities $S_{ij}$ defined in Eq.~\eqref{S_def} are real and anti-symmetric.

Due to the second term in Eq.~\eqref{M_off_diag}, the matrix element of $\Delta h$ depends on the PM-DHF potential and thereby on the yet to be determined PM-DHF core orbitals. Therefore, Eq.~\eqref{eig_eqn} is nonlinear and needs to be iterated until convergence. The iteration of Eq.~\eqref{eig_eqn} generally does not  suffer from the oscillating convergence behaviour as the finite-difference method. The change in the results from one iteration to another oscillates for the first few iterations but quickly decreases in a monotonous fashion. The price to be paid for this well-behaved convergence pattern is the need to pre-compute a large number of matrix elements of the form $g_{\bar{j}k\bar{k}j}-g_{\bar{j}\bar{k}kj}$ required in forming the $V'_{\rm HF}$ term in Eq.~\eqref{eig_eqn}.

Note also that since the $\bs{M}$ matrices corresponding to $\psi'_{\bar i}$ and $\psi'_i$ are related by swapping $j\leftrightarrow \bar{j}$ in Eq.~\eqref{M_def}, there is no need to diagonalize them separately. Instead, we form the $\bs{M}$ matrix only for negative values of $\kappa_i=-1,-2,\dots$. Each such matrix then has $4N$ eigenvectors, $2N$ of which correspond to the negative and positive energy orbitals $\psi'_i$ while the other $2N$ give the expansion for the orbitals $\psi'_{\bar i}$. We ensure the correct assignment of eigenvectors to orbitals by exploiting the fact that $\chi_{ii}\sim O(1)$, $\chi_{ij}\sim O(\eta^2)$ for $j\neq i$, and $\chi_{i\bar{j}}\sim O(\eta)$, in accordance with the results from perturbation theory. 

We now discuss the second method where, to avoid iterations in determining the PM-DHF core orbitals, one can also pre-compute them using the finite-difference solution of PM-DHF equations, see Sec.~\ref{Dalgarno_Lewis}. This is the strategy used earlier for basis set generation in the context of Breit interaction~\cite{derevianko_2001_Breit_NeutronSkin}. Then the required matrix elements of $\Delta h$, Eq.~\eqref{Eq:DeltaH}, can be computed immediately and the diagonalization proceeds in a single step. Comparing the PM-DHF core and low-lying excited orbitals from the finite-difference and basis-set solutions provides a valuable test of the accuracy. 
 
In both approaches, one has to be mindful of the smallness of the parameter $\eta\sim 10^{-15}$, which is comparable to the accuracy of double precision operations. Care should be taken when diagonalizing the matrix $\bs{M}$ to avoid numerical truncation errors. This issue may be effectively dealt with by using a multiple-precision diagonalization algorithm. In our numerical computations, we modified the routines {\fontfamily{qcr}\selectfont tred2} and {\fontfamily{qcr}\selectfont tqli} presented in Ref.~\cite{press1989numerical} to perform quadruple (128 bits) precision diagonalization and used these upgraded routines to diagonalize the matrices $\bs{M}$.

An alternative to matrix diagonalization is a perturbative approach that uses the smallness of parameter $\eta$, see Sec.~\ref{Perturbative matrix method}. However, the non-perturbative method described in this subsection is more general and is more accurate in the case of accidental degeneracies in the pseudo-spectra of $h_0$ between orbitals with the same total angular momentum but of opposite parities (see Sec.~\ref{Sec:NumSta} below for further discussions).  

We used the matrix diagonalization method discussed in this subsection to generate for \Cs\ a PM basis of total angular momenta ranging from $1/2$ to $13/2$ (one set for each method). The PP set used to expand the PM orbitals are B-splines obtained using the dual-kinetic-balance method~\cite{BELOY2008}. Each set of the PP partial waves with $\kappa_i\in\{\pm 1,\dots,\pm 7\}$ contains $N=40$ positive-energy orbitals. The cavity radius is chosen to be 50 a.u.~and computations were performed on a nonuniform grid of 500 points with 40 points inside the nucleus.

The PM core orbitals are read in from the finite-difference calculation and the PNC-DHF potential $V'_{\rm HF}-V_{\rm HF}$ is computed with these core orbitals. The rest of the PM basis is obtained by diagonalizing the matrices $M$ corresponding to $\kappa_i=-1,-2,\dots,-7$. The lowest order $6S_{1/2}-7S_{1/2}$ PNC frozen-core and core-perturbed amplitudes for Cs computed using the so-obtained PM-DHF valence orbitals $\psi'_{6s_{1/2}}$ and $\psi'_{6s_{1/2}}$ are, respectively
\begin{equation}\label{mat_dia_res}
\begin{aligned}
    E^{\rm fc}_{\rm PV} &= 0.73949\times10^{-11}\im|e|a_0(Q_W/N)\,,\\
    E^{\rm cp}_{\rm PV} &= 0.92701\times10^{-11}\im|e|a_0(Q_W/N)\,.
\end{aligned}
\end{equation} 
The differences between these basis-set values and the finite-difference results~\eqref{fin_dif_fro_cor_res} and \eqref{fin_dif_res} are at the level of 0.001\%. This numerical error is adequate for our goals.

\subsection{Parity-mixed basis set construction: Perturbative matrix method}\label{Perturbative matrix method}
The need for an iterative scheme and the numerical difficulty associated with the  smallness of the PNC matrix elements may be avoided entirely if we adopt \textit{ab initio} a form of expansion for the PM orbitals $\psi'_i$, Eq.~\eqref{opp_par_comb}, in accordance with perturbation theory. To the first order in $\eta$, perturbation theory tells us that 
\begin{equation}\label{conven_perb}
    \psi'_i = \psi^P_i + \sum_{\bar{j}}\frac{\bra{\bar{j}}\Delta h\ket{i}}{\varepsilon_i-\varepsilon_{\bar{j}}}\psi^P_{\bar{j}}\,,
\end{equation}
where the sum runs over all PP orbitals $\psi^P_{\bar{j}}$ with the same total angular momentum but opposite parity to $\psi^P_i$. More explicitly, Eq.~\eqref{conven_perb} has the form
\begin{equation}\label{conven_perb_exp}
\begin{aligned}
    \psi'_{n_ij_im_i} &= \psi^P_{n_i\ell_ij_im_i} \\
    &+ \sum_{n_{\bar{j}}}\frac{\bra{n_{\bar{j}}\bar{\ell}_ij_im_i}\Delta h\ket{n_i\ell_ij_im_i}}{\varepsilon_{n_i\ell_ij_i}-\varepsilon_{n_{\bar{j}}\bar{\ell}_ij_i}}\psi^P_{n_{\bar{j}}\bar{\ell}_ij_im_i}\,,
\end{aligned}
\end{equation}
where, again, the index $\bar{\ell}_i=\ell_i\pm1$ indicates that terms in the sum over $n_{\bar{j}}$ have opposite parities to $\psi^P_{n_i\ell_ij_im_i}$.

If the PM-DHF potential $V'_{\rm HF}$ has been constructed beforehand, e.g., by solving the finite difference Eqs.~\eqref{DL_eqns} for the PM core orbitals then Eq.~\eqref{conven_perb} may be used to directly compute the opposite-parity admixtures (the sum) for all PM excited orbitals. In contrast, if the PM core orbitals and the PM-DHF potential $V'_{\rm HF}$ are not known beforehand, the matrix method developed in Sec.~\ref{Matrix diagonalization method} may be used to solve for these orbitals as follows.

It is clear from Eq.~\eqref{conven_perb} that in a perturbative approach, the expansion coefficients $\chi_{ij}$ and $\chi_{i\bar{j}}$ in Eq.~\eqref{psi'_expansion} have the form
\begin{equation}\label{restrict_coeff}
	\begin{matrix}
		\chi_{ij} = \delta_{ij}\,, & \chi_{i\bar{j}} = \eta\gamma_{i\bar{j}}\,,
	\end{matrix}
\end{equation}
which makes it explicit that in the limit where $\eta\rightarrow 0$, $\psi'_i\rightarrow\psi^P_i$. Setting $\chi_{ii}=1$ guarantees that $\psi'_i$ is normalized up to $O(\eta^2)$. Factoring out the imaginary unit from the PNC corrections also makes sure that $\gamma_{i\bar{j}}$ are real and of order 1. 
% and $\varepsilon'_i\rightarrow \varepsilon_i$

Substituting the coefficients $\chi_{ij}$ and $\chi_{i\bar{j}}$ in Eqs.~\eqref{restrict_coeff} into Eq.~\eqref{ei2}, one obtains
\begin{equation}\label{perb_mat_eqn}
    \bra{\bar{j}}\Delta h\ket{i} = \im\eta(\varepsilon_i-\varepsilon_{\bar{j}})\gamma_{i\bar{j}}\,,
\end{equation}
which is the matrix equivalence of Eq.~\eqref{conven_perb}. We now need to solve Eq.~\eqref{perb_mat_eqn} for the unknown coefficients $\gamma_{i\bar{j}}$. For this purpose, we need to express the matrix element $\bra{\bar{j}}\Delta h\ket{i}$ in terms of the coefficients $\gamma_{i\bar{j}}$. Substituting Eqs.~\eqref{restrict_coeff} into Eq.~\eqref{M_off_diag} and replacing $j$ with $i$ therein, we find
\begin{equation}\label{h0'_mel}
	\begin{aligned}
		\bra{\bar{j}}\Delta h\ket{i}&= \im\eta\left[ S_{\bar{j}i}-\sum_{a\bar{k}}\gamma_{a\bar{k}}(g_{\bar{j}a\bar{k}i}-g_{\bar{j}\bar{k}ai})\right]\,,
	\end{aligned}
\end{equation}
where the summation runs over all PP core orbitals $\psi^P_a$ and all PP orbitals $\psi^P_{\bar k}$ which have the same total angular momentum but opposite parity to $\psi^P_a$

Substituting Eq.~\eqref{h0'_mel} into Eq.~\eqref{perb_mat_eqn}, one obtains
\begin{equation}\label{Eq_linearized}
    S_{\bar{j}i} -\sum_{a\bar{k}}\gamma_{a\bar{k}}(g_{\bar{j}a\bar{k}i}-g_{\bar{j}\bar{k}ai}) = (\varepsilon_i-\varepsilon_{\bar{j}})\gamma_{i\bar{j}}\,.
\end{equation}
Remember that in Eq.~\eqref{Eq_linearized}, the orbitals $\bar{j}$ have the same total angular momentum but opposite parity to the orbital $i$ whereas the orbitals $\bar{k}$ have the same total angular momentum but opposite parity to the orbital $a$. Equation~\eqref{Eq_linearized} allows us to solve for the PNC mixing coefficients $\gamma_{i\bar{j}}$. It is the matrix version of the finite-difference equations \eqref{DL_eqns}. In contrast with Eq.~\eqref{eig_eqn}, it is independent of the small parameter $\eta$ so is not subject to the issue with numerical inaccuracy as was the method described in Sec.~\ref{Matrix diagonalization method}. 

Let us consider the case where $i=b$, i.e., a core orbital. Denote by $N_c$ the number of core orbitals. We may then arrange all the coefficients $\gamma_{b\bar{j}}$ into a vector $\boldsymbol{\gamma}_b$ of length $2NN_c$, all the quantities $S_{\bar{j}b}$ into a vector $\bf S_b$ of length $2NN_c$, all the quantities $\varepsilon_b-\varepsilon_{\bar{j}}$ into a diagonal matrix $\Delta\varepsilon_b$ of size $2NN_c\times 2NN_c$ and all the quantities $g_{\bar{j}a\bar{k}b}-g_{\bar{j}\bar{k}ab}$ into a matrix $\bs{G}_b$ of size $2NN_c\times 2NN_c$. As a result, Eq.~\eqref{Eq_linearized} may be written in a more suggestive form as
\begin{equation}\label{Core_Perb}
    {\bf S}_b - \bs{G}_b\boldsymbol{\gamma}_b =\Delta\varepsilon_b\boldsymbol{\gamma}_b\,,
\end{equation}
whose solution reads
\begin{equation}\label{eig_soln}
    \boldsymbol{\gamma}_b = (\Delta\varepsilon_b+\bs{G}_b)^{-1}{\bf S}_b\,.
\end{equation}
Equation~\eqref{eig_soln} allows us to obtain the mixing coefficients $\gamma_{b\bar{j}}$ for all core orbitals. We point out that Eq.~\eqref{eig_soln} is linear so there is no need for an iterative scheme as with the methods discussed in Secs.~\ref{Dalgarno_Lewis} and~\ref{Matrix diagonalization method}.

After solving for the PNC mixing coefficients $\gamma_{b\bar{j}}$ of all $N_c$ core orbitals, we again use Eq.~\eqref{Eq_linearized} to solve for the mixing coefficients of all unoccupied orbitals $\psi'_m$, obtaining
\begin{equation}\label{soln_d_valence}
	\gamma_{m\bar{j}}=\frac{S_{\bar{j}m}-\sum_{a\bar{j}}\gamma_{a\bar{k}}(g_{\bar{j}a\bar{k}m}-g_{\bar{j}\bar{k}am})}{\varepsilon_m-\varepsilon_{\bar{j}}}\,.
\end{equation}
In this form, Eq.~\eqref{soln_d_valence} clearly demonstrates the perturbative nature of the current approach. As a result, during computation, one should check that accidental degeneracy does not happen, or in other words, that the coefficients $|\eta\gamma_{m\bar{j}}| \ll 1$. If such event does occur, the more general method described in Sec.~\ref{Matrix diagonalization method} should be used instead.

We used the perturbative matrix method discussed in this subsection to generate for \Cs\ a PM basis of total angular momenta ranging from $1/2$ to $13/2$. The PP set used to expand the PM orbitals are the same as that used in Sec.~\ref{Matrix diagonalization method}. The lowest order $6S_{1/2}-7S_{1/2}$ PNC frozen-core and core-perturbed amplitudes for Cs computed using the so-obtained PM-DHF valence orbitals $\psi'_{6s_{1/2}}$ and $\psi'_{6s_{1/2}}$ are, respectively
\begin{equation}\label{mat_perb_res}
\begin{aligned}
    E^{\rm fc}_{\rm PV} &= 0.73947\times10^{-11}\im|e|a_0(Q_W/N)\,,\\
    E^{\rm cp}_{\rm PV} &= 0.92697\times10^{-11}\im|e|a_0(Q_W/N)\,.
\end{aligned}
\end{equation} 
The small differences between the results \eqref{mat_dia_res} and \eqref{mat_perb_res} of the two matrix methods may be attributed to nonlinear $O(\eta^2)$ terms, which, although small, may propagate through the computation. At the level of 0.004\%, these numerical differences are acceptable for our goals as we ultimately aim at 0.2\% overall accuracy in the PNC amplitude.

\subsection{Numerical stability of parity-mixed basis sets}\label{Sec:NumSta}

In the previous sections, we have presented different methods through which basis sets of PM single-electron orbitals may be obtained. Before discussing the application of these basis sets in MBPT and CC calculations, we pause here to make a few remarks regarding their numerical stability, specifically with respect to the small parameter $\eta$. 

In the finite difference and perturbative matrix methods, a PM single-electron orbital is expanded into two components of opposite parities:  a real part being independent of $\eta$ and an imaginary part having a linear dependence on $\eta$. Furthermore, as was shown in Secs.~\ref{Dalgarno_Lewis} and~\ref{Perturbative matrix method}, the factor $\eta$ may be completely separated from the imaginary part, allowing one to reliably compute this component. At the DHF level, the PNC transition amplitude $E_{\rm PV}$ obtained using the resulting PM orbitals reads
\begin{align}\label{how_EPV}
    E_{\rm PV}&=\bra{\psi'_{6s_{1/2}}}D_z\ket{\psi'_{7s_{1/2}}}\nonumber\\
    &=\im \eta(\bra{\psi_{6s_{1/2}}}D_z\ket{\bar{\psi}_{7s_{1/2}}}-\bra{\bar{\psi}_{6s_{1/2}}}D_z\ket{\psi_{7s_{1/2}}})\,,
\end{align}
which shows that $E_{\rm PV}$ depends linearly on $\eta$.

\begin{figure}[htb]
    \centering
    \includegraphics[scale=0.3]{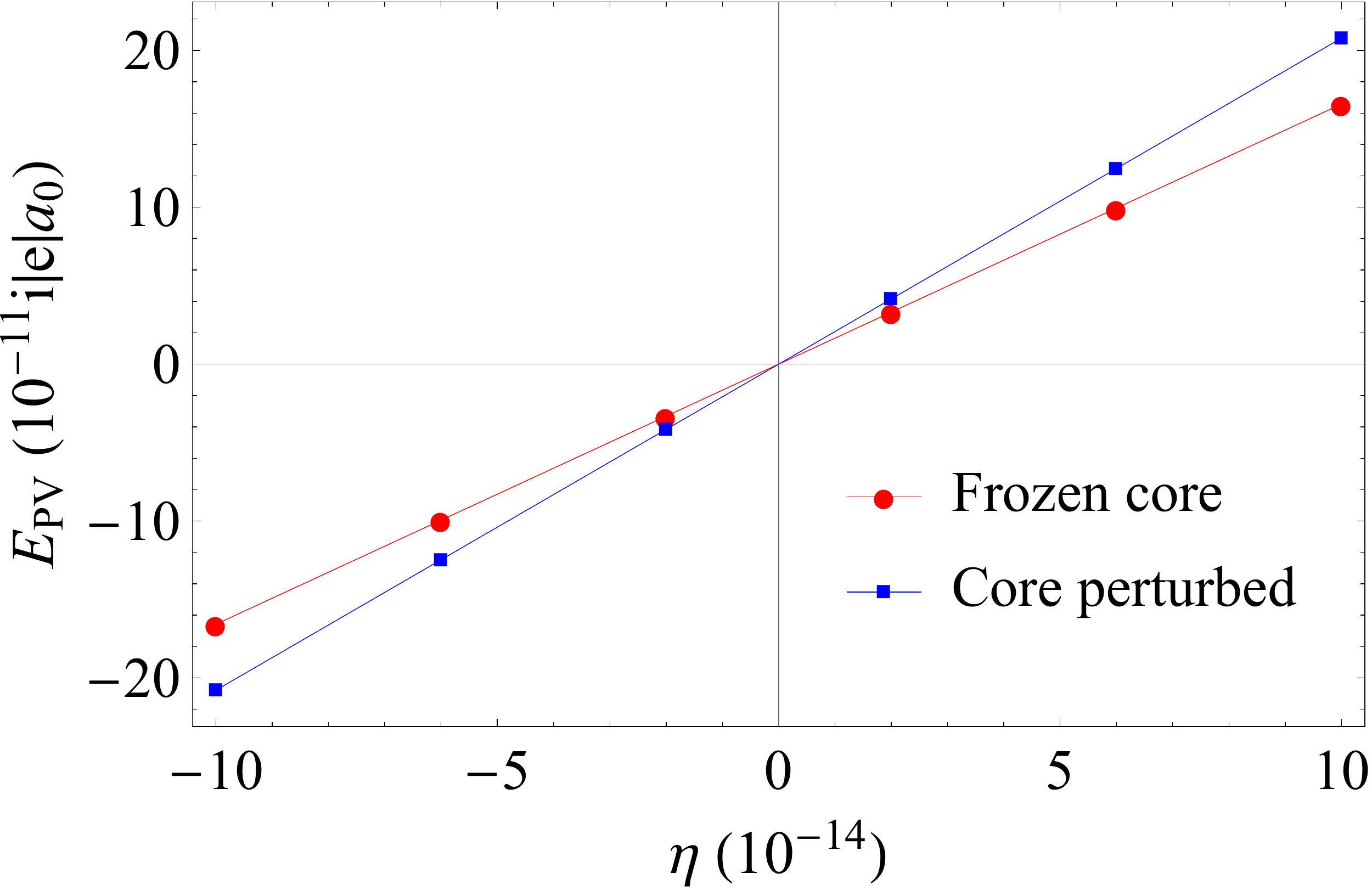}
    \caption{(Color online) The dependence on the dimensionless parameter $\eta\equiv G_FQ_W/(2\sqrt{2}a_0^2)$ of the $6S_{1/2}-7S_{1/2}$ PNC transition amplitudes (in both frozen core and core perturbed approximations) in \Cs\ calculated using the exact matrix diagonalization method described in Sec.~\ref{Matrix diagonalization method}. The lines being straight demonstrate that the effects that are nonlinear in $\eta$ do not show when computing $E_{\rm PV}$.}
    \label{EPV_Linear}
\end{figure}

In contrast, if the exact matrix diagonalization method is used, the resulting PM single-electron orbitals contain, in principle, nonlinear dependence on $\eta$. As remarked in Sec.~\ref{Matrix diagonalization method}, this is due to the need of solving the nonlinear eigenvalue Eq.~\eqref{eig_eqn}. As a result, the PNC transition amplitude $E_{\rm PV}$, computed as in Eq.~\eqref{how_EPV} will also contain contributions nonlinear in $\eta$. However, these nonlinear contributions are not manifest at the level of accuracy we are interested in, as may be observed from Fig.~\ref{EPV_Linear}, which shows the linear dependence on $\eta$ of the PNC transition amplitudes $E^{\rm fc}_{\rm PV}$ and $E^{\rm cp}_{\rm PV}$ calculated using the exact matrix diagonalization method.

Similarly, it may be argued that when PM single-electron orbitals are used in the MBPT and CC computations, terms that are $O(\eta^2)$ or higher do not contribute numerically. This justifies our direct upgrade of the conventional PP-MBPT and PP-CC formalism to the PM ones without having first to linearize their equations in terms of $\eta$. At the desired level of numerical accuracy $\ll 0.2\%$, contributions that are $O(\eta^2)$ or higher simply do not show up.

We end this section by elaborating on the advantage of the exact matrix diagonalization method in the case of accidental (near) degeneracy between states with the same angular momentum but opposite parities. We stress that this degeneracy can appear as  an artefact of using finite basis set of  orbitals (pseudo-spectrum).
%instead of those in the physical one. 
Two orbitals $\psi_1$ and $\psi_2$ are considered to be nearly degenerate if the perturbation theory convergence criterion,   $|\bra{\psi_1}h_W\ket{\psi_2}/(\varepsilon_1-\varepsilon_2)| \ll 1$, fails. This problem may be avoided by varying the parameters of the basis set, such as the radius of the cavity, so as to make all quantities of the $|\bra{\psi_1}h_W\ket{\psi_2}/(\varepsilon_1-\varepsilon_2)|$ form to be much smaller than 1. We test our numerical sets for these accidental degeneracies before applying the perturbative approach.

Alternatively, this tuning of the basis set parameters may be avoided by using matrix diagonalization: quantities of the form $|\bra{\psi_1}h_W\ket{\psi_2}/(\varepsilon_1-\varepsilon_2)|$ do not arise in this method. It is worth noting also that in this case, the lifting of degeneracy by $h_W$ is $O(\eta)$. For example, consider again two states $\psi_1$ and $\psi_2$ of the same total angular momentum, opposite parities, and energies $\varepsilon_1\approx\varepsilon_2=\varepsilon$. To find the energy corrections due to the perturbation $h_W$, one solves the secular equation for the perturbed energy $\varepsilon'$
\begin{equation}
    {\rm det}\begin{pmatrix}
    \varepsilon-\varepsilon' && \bra{\psi_1}h_W\ket{\psi_2}\\
    \bra{\psi_2}h_W\ket{\psi_2}&&\varepsilon-\varepsilon'\end{pmatrix}=0\,,
\end{equation}
obtaining
\begin{equation}
    \varepsilon'=\varepsilon\pm |\bra{\psi_1}h_W\ket{\psi_2}|\,,
\end{equation}
which shows that the energy corrections are $O(\eta)$ for degenerate states. Note that in this case,
strictly speaking, one also needs to include the natural decay widths $\Gamma_i$ to the energy levels, $ \varepsilon_i \rightarrow \varepsilon_i - \im \Gamma_i/2$ which can lift the degeneracy and requires further modifications to the code. 

\section{Matrix elements in the parity-mixed basis}\label{Sec:MatElsPMBasis}
Now with the PM basis constructed, we go back to the MBPT formalism of Sec.~\ref{Theory}.
The basic building blocks of MBPT expressions are the matrix elements of one-body (e.g., the electric dipole) and 
two-body (e.g., Coulomb interaction) operators in the PM basis, $\psi'_i =   \psi_i +\im \eta\bar{\psi}_i$. Due to the smallness of the parameter $\eta$, we may linearize the resulting expressions in $\eta$. Then any matrix element of an operator of definite parity splits into a part involving only the PP orbitals $\psi_i$ and a correction that involves opposite parity admixtures $\bar{\psi}_i$ (PNC correction). The former is already implemented in traditional MBPT codes. The latter may be readily added to these codes by modifying parity selection rules and using the radial components of $\bar{\psi}_i$. 

Furthermore, we show that the matrix elements of any operator of definite parity in the PM basis is either purely real- or imaginary-valued. With our phase convention for the PM radial components~\eqref{PNC_DHF_para}, PP parts are always real, while the PNC corrections are always imaginary. We will exploit this fact to derive useful symmetries of the reduced matrix elements of the one- and two-body operators. These symmetries will help significantly reduce the amount of computation and storage needed in MBPT calculations.

\subsection{Angular reduction of matrix elements}
Let us begin by discussing the angular reduction of the PM matrix elements of one- and two-body operators. The particular operators of interest here are of course the electric dipole operator and the inter-electron Coulomb interaction. 

Without loss of generality, we assume that the operators in question are Hermitian and can be represented as components of irreducible tensor operators. We also observe that the PM orbitals are eigen-states of the total angular momentum operators $J_z$ and $\mb{J}^2$. As a result, the Wigner-Eckart theorem applies to the matrix elements of the one- and two-body operators with respect to these PM orbitals.
Moreover, since the weak interaction is a pseudo-scalar, a PM orbital has the very same total angular momentum as its PP counterpart. 

As a result, the angular reduction of a one-body matrix element $t'_{ij}$ has the same form as that of the PP $t_{ij}$. More explicitly, for the case where $t=z$, the Wigner-Eckart theorem states
\begin{equation}\label{z'}
	\begin{aligned}
		z'_{kl}=(-1)^{j_k-m_k}\begin{pmatrix}
			j_k & 1 & j_l\\
			-m_k & 0 &m_l
		\end{pmatrix}\rmel{k}{z}{l}'\,,
	\end{aligned}
\end{equation}
where all the information about mixing parities is contained in the reduced matrix element $\rmel{k}{z}{l}'$.

Similarly, the angular reduction of the PM Coulomb integrals $g'_{ijkl}$ is the same as that for the PP $g_{ijkl}$, namely %\cite{PalSafJoh07}
\begin{equation}\label{g_ijkl_PNC}
    \begin{aligned}
        g'_{ijkl}&=\sum_LJ_L(ijkl)X'_L(ijkl)\,,\\
        J_L(ijkl)&\equiv\sum_{M}(-1)^{j_i-m_i+j_j-m_j}\\
        &\times\begin{pmatrix}
        j_i & L & j_k \\
        -m_i & -M & m_k
        \end{pmatrix}\begin{pmatrix}
        j_j & L & j_l \\
        -m_j & M & m_l
        \end{pmatrix}\,,
    \end{aligned}
\end{equation}
where all the information about mixing parities is contained in the reduced matrix element $X'_L(ijkl)$.

We may write the PM reduced matrix elements of a one-body operator $t$ as (since $\psi'_i =   \psi_i +i\eta\bar{\psi}_i$)
\begin{subequations}
    \begin{align}\label{zmel_PNC}
		\rmel{k}{t}{l}'&= \rmel{k}{t}{l} + \im \eta \rmel{k}{t}{l}''  \,,\\
		\rmel{k}{t}{l}''&\equiv  \rmel{k}{t}{\bar{l} } - \rmel{\bar{k}}{t}{ l}\,,
    \end{align}
\end{subequations}
where have we dropped $O(\eta^2)$ terms. Explicitly, for an electric-dipole operator $z$, relevant to computing PNC amplitudes, the three reduced matrix elements read 
\begin{equation}\label{PP_redux_z}
\begin{aligned}
\rmel{k}{z}{l} &= \rmel{\kappa_k}{C_1}{\kappa_l}\int\left(P_kP_l+Q_kQ_l\right)rdr\,, \\
\rmel{k}{z}{\bar{l} } &= \rmel{\kappa_k}{C_1}{\kappa_{\bar{l}}}\int\left(P_k\bar{P}_l+Q_k\bar{Q}_l\right)rdr\,,\\
\rmel{\bar{k}}{z}{l}&=  \rmel{\kappa_{\bar{k}}}{C_1}{\kappa_l}\int\left(\bar{P}_kP_l+\bar{Q}_kQ_l\right)rdr\,.
\end{aligned}
\end{equation}
Here $\rmel{k}{z}{l}$  is the PP contribution, and the $\rmel{k}{z}{\bar{l}}$ and $\rmel{\bar{k}}{z}{l}$ contributions are due to
the opposite parity admixtures as indicated by the large and small radial components with overhead bars, c.f.~Eq.~\eqref{PNC_DHF_para}.

The parity selection rules are encoded into the reduced matrix elements of the normalized spherical harmonic via
$\rmel{\kappa_k}{C_1}{\kappa_l} \propto{\rm mod}_2( \ell_k + \ell_l)$, i.e. $\ell_k + \ell_l$ must be odd. Similar selection rules apply to the P-odd corrections, e.g., $\rmel{\kappa_{\bar{k}}}{C_1}{\kappa_l}\propto {\rm mod}_2( \ell_k + \ell_l+1)$, since $\ell_{\bar{k}} = \ell_{k} \pm 1$, i.e. $\ell_k + \ell_l$ must be even.  Note that for two fixed PM orbitals, these selection rules cannot be satisfied simultaneously, thereby, the matrix element is either pure real- or imaginary-valued. With our phase convention for the PM radial components, the PP part is always real, while the PNC correction is always imaginary.
The above statements can be easily generalized to any irreducible tensor operator of definite parity.

Similar considerations apply to the reduced Coulomb matrix element:
\begin{subequations}
    \begin{align}
        X'_L(ijkl) &= X_L(ijkl) + \im\eta X''_L(ijkl)\label{X_L^PNC}\,,\\
        X''_L(ijkl)&\equiv X_L(ij\bar{k}l)+ X_L(ijk\bar{l})\nonumber\\
        &- X_L(\bar{i}jkl) - X_L(i\bar{j}kl)\,.\label{X''}
\end{align}
\end{subequations}
Here, the quantity $X_L(ijkl)$ is expressed in terms of the reduced matrix element of the normalized spherical harmonic  $C_L(\hat{\bf r})$ and the Slater integral $R_L(ijkl)$:
\begin{equation}\label{PP_redux_X}
     X_L(ijkl) = (-1)^L\rmel{\kappa_i}{C_L}{\kappa_k}\rmel{\kappa_j}{C_L}{\kappa_l}R_L(ijkl)\,.
\end{equation}
The parity selection rules for $X_L(ijkl)$ are $(-1)^{\ell_i + L + \ell_k}=+1$ and $(-1)^{\ell_j + L + \ell_l}=+1$.  

The quantities in Eq.~\eqref{X''} are defined similarly. For example, 
\begin{equation}
	X_L(\bar{i}jkl) = (-1)^L\rmel{\kappa_{\bar{i}}}{C_L}{\kappa_k}\rmel{\kappa_j}{C_L}{\kappa_l}R_L(\bar{i}jkl)\,,
\end{equation}
where the index $\bar{i}$ in $R_L(\bar{i}jkl)$ means that we use the radial functions $\bar{P}_i$ and $\bar{Q}_i$ as defined in Eq.~\eqref{psi_bar}.

The parity selection rules for the various terms in Eqs.~\eqref{X_L^PNC} and \eqref{X''} are also clear. If $\ell_i+L+\ell_k$ and $\ell_j+L+\ell_l$ are both even then $X'_L(ijkl)=X_L(ijkl)$ which is purely real whereas if they are both odd then $X'_L(ijkl)=0$. If $\ell_i+L+\ell_k$ is odd but $\ell_j+L+\ell_l$ is even then $X'_L(ijkl)=\im\eta X''_L(ijkl)$ is purely imaginary and $X''_L(ijkl)$ is given by the first two terms in Eq.~\eqref{X''}. On the other hand, if $\ell_i+L+\ell_k$ is even but $\ell_j+L+\ell_l$ is odd then $X''_L(ijkl)$ is given by the last two terms in Eq.~\eqref{X''}. Translating to $g'_{ijkl}$, these rules mean that if $\ell_i+\ell_j+\ell_k+\ell_l$ is even then $g'_{ijkl}$ is real and equals its PP counterpart $g_{ijkl}$, whereas if $\ell_i+\ell_j+\ell_k+\ell_l$ is odd then $g'_{ijkl}$ is purely imaginary. These observations will prove useful in the formulation of the PM-CC formalism, Sec~\ref{Sec:PMCC}.

Another frequently occurring matrix element is that of the anti-symmetrized Coulomb interaction, $\tilde{g}'_{ijkl} \equiv {g}'_{ijkl}-{g}'_{ijlk}$, which can be brought in the angular-diagram form identical to that of ${g}'_{ijkl}$, c.f. Eq.~\eqref{g_ijkl_PNC},
\begin{equation}\label{g_tilde'}
    \begin{aligned}
        \tilde{g}'_{ijkl}\equiv{g}'_{ijkl}-{g}'_{ijlk}&=\sum_{L}J_L(ijkl)Z'_L(ijkl)\,,
    \end{aligned}
\end{equation}
where the reduced matrix element is given by
\begin{subequations}
\begin{align}
        Z'_L(ijkl) &= Z_L(ijkl) + \im\eta Z''_L(ijkl)\label{Z'}\\
        Z''_L(ijkl)&\equiv Z_L(ij\bar{k}l)+Z_L(ijk\bar{l})\nonumber\\
        &- Z_L(\bar{i}jkl) - Z_L(i\bar{j}kl)\,.\label{Z''}
\end{align}
\end{subequations}

In these equations, $Z_L(ijkl)$ may be expressed in terms of $X_L(ijkl)$ via
	\begin{align}\label{PP_redux_Z}
		Z_L(ijkl) &= X_L(ijkl) \nonumber\\
				  &+ \sum_{L'} (2L+1)\begin{Bmatrix}
					j_k & j_i & L \\
					j_l & j_j & L'
		\end{Bmatrix}X_{L'}(ijlk)\,,
	\end{align}
and the other quantities are defined similarly. Again, the overhead bars in Eq.~\eqref{Z''} signify the use of the $P$-odd radial functions $\bar{P}_i$ and $\bar{Q}_i$ as defined in Eq.~\eqref{psi_bar}. The parity selection rules for $g'_{ijkl}$ also apply to $\tilde{g}'_{ijkl}$, namely, $\tilde{g}'_{ijkl}$ is real and equals its PP counterpart if $\ell_i+\ell_j+\ell_k+\ell_l$ is even whereas $\tilde{g}'_{ijkl}$ is purely imaginary if $\ell_i+\ell_j+\ell_k+\ell_l$ is odd.

To reiterate, we observe that the PM matrix elements, Eqs.~\eqref{z'} and \eqref{g_tilde'}, split into real PP parts and purely imaginary PNC parts. Due to the small coefficient $\eta$, the imaginary parts are many orders of magnitude smaller than the real parts. This, however, does not give rise to a problem with numerical accuracy due to truncation as mentioned in Sec.~\ref{Matrix diagonalization method}. In fact, if the MBPT code is modified to use complex instead of real numbers and the PP and PNC parts are stored separately then algebraic operations will always involve adding terms of the same order of magnitude. In Sec.~\ref{Many-body calculations of PNC amplitude}, we shall present how such a procedure is carried out with the example of the random-phase approximation (RPA).

\subsection{Symmetries of reduced matrix elements}
Before starting with the discussion of RPA, however, let us present the symmetries of the reduced matrix elements $\rmel{k}{z}{l}'$, $X'_L(ijkl)$ and $Z'_L(ijkl)$ with respect to the exchange of their indices. In a conventional MBPT formalism which uses PP single-electron orbitals, the corresponding symmetries of the PP reduced matrix elements are exploited to a great extent to significantly reduce the amount of computation and storage needed. We will show that similar symmetries are also available for a MBPT scheme using PM orbitals so the same economy may be achieved.

We begin with the matrix elements of a one-body operator. For our purpose, we concentrate on the electric dipole operator $z$. Using the definitions \eqref{PP_redux_z},~\eqref{PP_redux_X}, and \eqref{PP_redux_Z} and the following property of the reduced matrix elements of the normalized spherical harmonics
\begin{equation}\label{C_L_symm}
    \rmel{\kappa_k}{C_L}{\kappa_l} = (-1)^{j_k-j_l}\rmel{\kappa_l}{C_L}{\kappa_k}\,,
\end{equation}
it may be verified that the PP reduced matrix elements of $z$ satisfy
\begin{equation}\label{57a}
    \rmel{k}{z}{l} = (-1)^{j_k-j_l}\rmel{l}{z}{k}\,.
\end{equation}
Next, by using Eqs.~\eqref{zmel_PNC} and \eqref{57a}, we find the following symmetry for the PM reduced matrix elements of the electric dipole moment
\begin{equation}\label{symm_z'}
    \rmel{k}{z}{l}' = (-1)^{j_k-j_l}[\rmel{l}{z}{k}']^*\,,
\end{equation}
where $*$ denotes complex conjugation.

We note that although we only considered the electric dipole operator $z$, the result presented above applies to any single-electron irreducible tensor operator of rank $k$, $T^{(k)}$, namely
\begin{equation}\label{symm_Tk}
    \rmel{k}{T^{(k)}}{l}' = (-1)^{j_k-j_l}[\rmel{l}{T^{(k)}}{k}']^*\,.
\end{equation}

We now turn to the reduced matrix elements of the inter-electron Coulomb interaction. We begin by presenting the familiar symmetries of the PP reduced matrix elements $X_L(ijkl)$ and $Z_L(ijkl)$. Although these results are not new, they serve as a convenient reference point for our discussion of the PM matrix elements. Using the definitions~\eqref{PP_redux_X} and \eqref{PP_redux_Z} and the property \eqref{C_L_symm}, one easily finds the following relations
\begin{subequations}
    \begin{align}
    X_L(ijkl) &= X_L(jilk)\,,\label{57b}\\
    X_L(ijkl) &= (-1)^{j_i-j_k}X_L(kjil)\,,\label{57c}\\
    X_L(ijkl) &= (-1)^{j_i+j_j+j_k+j_l}X_L(klij)\,,\label{57d}\\
    Z_L(ijkl) &= Z_L(jilk)\,,\label{57e}\\
    Z_L(ijkl) &= (-1)^{j_i+j_j+j_k+j_l}Z_L(klij)\,,\label{57f}\\
    Z_L(ijkl) &= [L]\sum_{L'}\left\{\begin{matrix}
        j_j && j_l && L \\
        j_i && j_k && L'
    \end{matrix}\right\}Z_{L'}(jikl)\,,\label{last_symm}
\end{align}
\end{subequations}
where $[L]\equiv 2L+1$ and $\left\{\begin{matrix}
        j_j && j_l && L \\
        j_i && j_k && L'
    \end{matrix}\right\}$ is the $6j$-symbol.

From the expansions \eqref{X_L^PNC} and \eqref{Z'} for $X'_L(ijkl)$ and $Z'_L(ijkl)$ and the properties \eqref{57b} and \eqref{57e}, one sees that simultaneously swapping $i\leftrightarrow j$ and $k\leftrightarrow l$ has no effect on the Coulomb reduced matrix elements, i.e.,
\begin{subequations}\label{symm_XZ'_jilk}
    \begin{align}
    X'_L(ijkl) &= X'_L(jilk)\,,\\
    Z'_L(ijkl) &= Z'_L(jilk)\,.
\end{align}
\end{subequations}

It may also be observed from Eqs.~\eqref{57d} and \eqref{57f} that swapping the pair $ij\leftrightarrow kl$ is equivalent to introducing the phase factor $(-1)^{j_i+j_j+j_k+j_l}$ to $X'_L(ijkl)$ and $Z'_L(ijkl)$ as well as switching the sign of $X''_L(ijkl)$ and $Z''_L(ijkl)$. As a result, we have
\begin{subequations}\label{symm_XZ'_klij}
    \begin{align}
    X'_L(ijkl) &= (-1)^{j_i+j_j+j_k+j_l}[X'_L(klij)]^*\,,\\
    Z'_L(ijkl) &= (-1)^{j_i+j_j+j_k+j_l}[Z'_L(klij)]^*\,.
\end{align}
\end{subequations}

Next, we present the PM equivalence of Eq.~\eqref{57c}. For this purpose, it is convenient to consider two separate cases. First, let us assume that the nominal parities of the orbitals $\psi'_i$ and $\psi'_k$ satisfy the condition $(-1)^{\ell_i+L+\ell_k}=1$. In this case, Eq.~\eqref{X_L^PNC} simplifies to
\begin{align}
        X'_L(ijkl) = X_L(ijkl) + \im\eta\left[X_L(ijk\bar{l})-X_L(i\bar{j}kl)\right]\,,
    \end{align}
which, when combined with Eq.~\eqref{57c}, gives
\begin{equation}\label{ik_even}
    X'_L(ijkl) = (-1)^{j_i-j_k}X'_L(kjil)\,,
\end{equation}
if $(-1)^{\ell_i+L+\ell_k}=1$. On the other hand, if $(-1)^{\ell_i+L+\ell_k}=-1$ then  Eq.~\eqref{X_L^PNC} simplifies to
\begin{equation}\label{63}
    X'_L(ijkl) = \im\eta\left[X_L(ij\bar{k}l) - X_L(\bar{i}jkl)\right]\,.
\end{equation}
It is clear from Eq.~\eqref{63} that in this case, swapping $i\leftrightarrow k$ introduces the factor $(-1)^{j_i-j_k}$ as well as a minus sign. Thus, we have
\begin{equation}\label{ik_odd}
    X'_L(ijkl) = -(-1)^{j_i-j_k}X'_L(kjil)\,.
\end{equation}
if $(-1)^{\ell_i+L+\ell_k}=-1$. We may combine Eqs.~\eqref{ik_even} and \eqref{ik_odd} into a single formula, writing
\begin{align}\label{symm_X'_kjil}
    X'_L(ijkl) = (-1)^{\ell_i+L+\ell_k}(-1)^{j_i-j_k}X'_L(kjil)\,,
\end{align}
which is the PM equivalence of Eq.~\eqref{57c}.

Finally, since the recoupling rule Eq.~\eqref{last_symm} involves only total angular momenta and no sign change, its PM version has the same form, i.e.,
\begin{equation}\label{last_symm_PM}
    Z'_L(ijkl) = [L]\sum_{L'}\left\{\begin{matrix}
        j_j && j_l && L \\
        j_i && j_k && L'
    \end{matrix}\right\}Z'_{L'}(jikl)\,.
\end{equation}

Equations \eqref{symm_z'},~\eqref{symm_XZ'_jilk},~\eqref{symm_XZ'_klij},~\eqref{symm_X'_kjil} and \eqref{last_symm_PM} represent the symmetries of the PM reduced matrix elements of the electric dipole and inter-electron Coulomb interaction operators with respect to permutations of the PM orbitals. They will be used extensively in the PM-MBPT as well as PM-CC calculations.

\section{Random-phase approximation for the parity non-conserving amplitude}\label{Many-body calculations of PNC amplitude}
In Sec.~\ref{PM basis functions} we presented several methods through which basis sets of PM orbitals may be constructed. The numerical accuracy of these basis sets was tested by computing the PNC amplitude between the PM-DHF valence states. Strictly speaking, this test only
involves two single-electron PM-DHF valence orbitals, $|6s_{1/2}'\rangle$ and $|7s_{1/2}'\rangle$. In Sec.~\ref{Sec:MatElsPMBasis} we discussed formulas for the matrix elements of one- and two-body operators, in particular the electric dipole and Coulomb operators, in terms of the PM-DHF bases. These matrix elements are needed in the MBPT paradigm to take into account the effects of inter-electron correlation on the PNC amplitude. In this section, we shall use these formulas to compute the second-order and RPA all-order correlation corrections to the matrix elements of the electric dipole operator.

The relevant second-order formula for PNC amplitude is given in Eq.~\eqref{dtwv} and it involves summations over the entire PM-DHF basis set.  Here, using this formula, we test 
the accuracy of our generated PM-DHF basis sets by computing PNC amplitude in \Cs\ in the well-established random-phase approximation (RPA)~\cite{BLUNDELL1987,Dzuba1985,DZUBA1989,Johnson1996}. %\APD{Add Dzuba's papers} 
RPA sums diagrams topologically similar to second-order Eq.~\eqref{dtwv} to all orders of MBPT. This not only tests the quality of PM-DHF basis sets, but importantly builds the foundation for the formulation of  
parity-mixed coupled-cluster (PM-CC) method, which systematically enables all-order summations of substantially larger classes of diagrams, see  Sec.~\ref{Sec:PMCC}.

For now, we focus on the RPA method. In this approximation, one first takes into account the second-order correction to the ``core-to-excited'' matrix elements $t'_{an}$ and $t'_{na}$ present in Eq.~\eqref{dtwv}. Denoted by $t'^{\rm RPA}_{an}$ and $t'^{\rm RPA}_{na}$ the RPA vertices, one finds that these quantities satisfy equations similar to Eq.~\eqref{dtwv}, namely
\begin{equation}\label{RPA_core}
	\begin{aligned}
		t'^{\rm RPA}_{an} &= t'_{an} + \sum_{bm}\frac{t'^{\rm RPA}_{bm}\tilde{g}'_{amnb}}{\varepsilon'_b-\varepsilon'_m-\omega}+ \sum_{bm}\frac{\tilde{g}'_{abnm}t'^{\rm RPA}_{mb}}{\varepsilon'_b-\varepsilon'_m+\omega}\,,\\
		t'^{\rm RPA}_{na} &= t'_{na} + \sum_{bm}\frac{t'^{\rm RPA}_{bm}\tilde{g}'_{nmab}}{\varepsilon'_b-\varepsilon'_m-\omega}+ \sum_{bm}\frac{\tilde{g}'_{nbam}t'^{\rm RPA}_{mb}}{\varepsilon'_b-\varepsilon'_m+\omega}\,,
	\end{aligned}
\end{equation}  
which will be solved by iteration to convergence. Once the RPA vertices
are obtained, the matrix elements between two valence orbitals $\psi'_v$ and $\psi'_w$  are given by 
\begin{equation}\label{RPA_valence}
	\begin{aligned}
		T'_{wv} = t'_{wv} + \sum_{an}\frac{t'^{\rm RPA}_{an}\tilde{g}'_{wnva}}{\varepsilon'_a-\varepsilon'_n-\omega}+ \sum_{an}\frac{\tilde{g}'_{wvna}t'^{\rm RPA}_{na}}{\varepsilon'_a-\varepsilon'_n+\omega}\,.
	\end{aligned}
\end{equation}
For computations of \Cs\ PNC amplitudes,  $t$ is the electric-dipole operator, and $v=6s$, $w=7s$.

\begin{figure}[htb]
    \centering
    \includegraphics[scale=0.3]{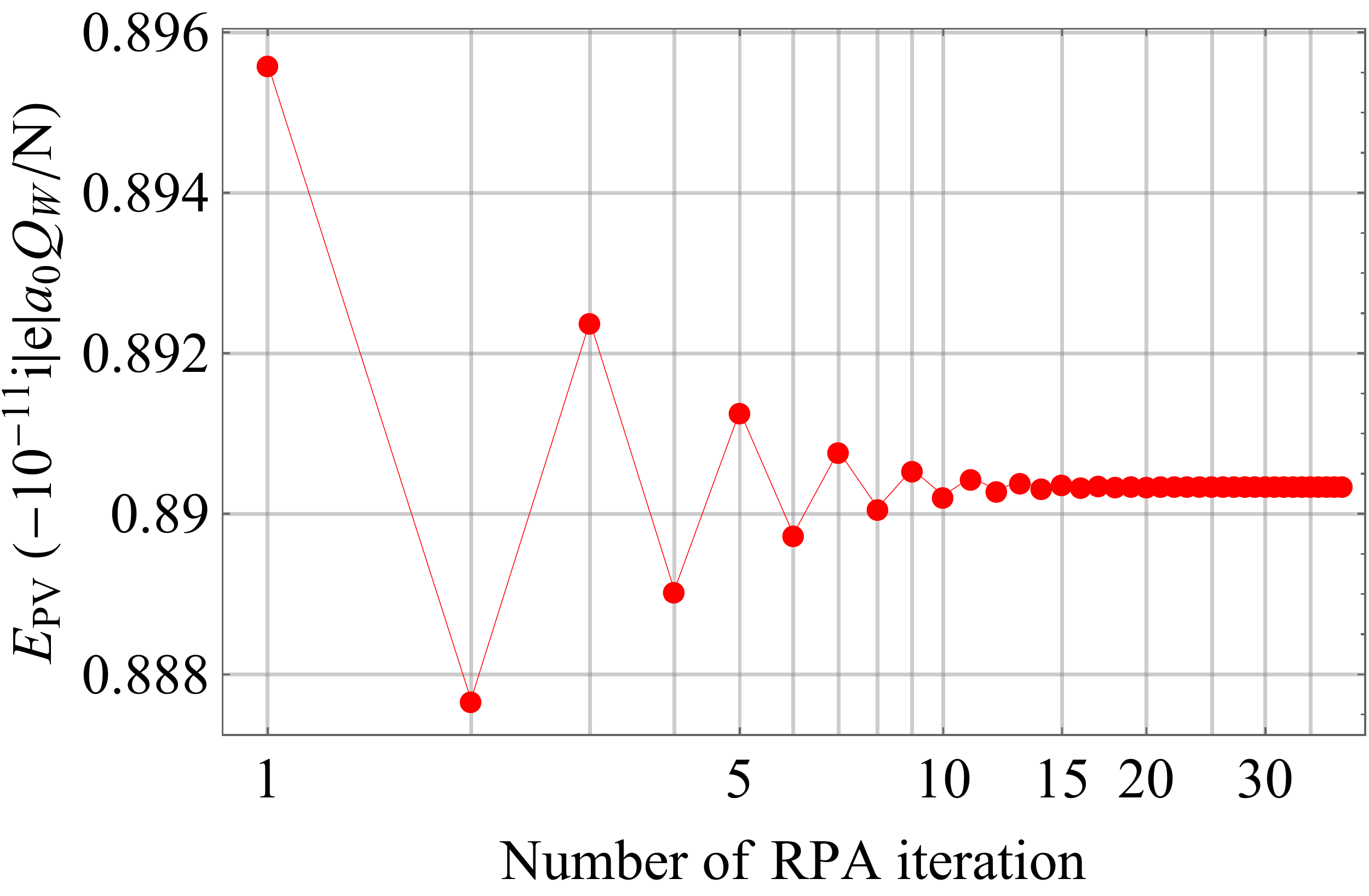}
    \caption{(Color online) The value of the $6S_{1/2}-7S_{1/2}$ PNC transition amplitude in \Cs\ as a function of the number of RPA iteration. Convergence occurs after 20 iterations at the level of fractional accuracy of $10^{-6}$.}
    \label{RPA_Conv}
\end{figure}

We used the PM-DHF basis sets of Sec.~\ref{PM basis functions} to compute the RPA correction to the $6S_{1/2}-7S_{1/2}$ PNC transition amplitude in Cs. The forms of the dipole matrix elements $z'_{wv}$ and the Coulomb matrix elements $\tilde{g}'_{wnva}$ and $\tilde{g}'_{wvna}$ needed for this computation were presented in Eqs.~\eqref{z'} and \eqref{g_tilde'}. The resulting value of the amplitude as a function of the number of RPA iteration is shown in Fig.~\ref{RPA_Conv} where the oscillatory behavior typical of an RPA calculation is clearly visible. The final value for the $6S_{1/2}-7S_{1/2}$ PNC amplitude is at 
\begin{equation}\label{RPA_res}
	E^{\rm RPA}_{\rm PV} = 0.89034\times10^{-11}\im|e|a_0(Q_W/N)\,.
\end{equation} 
This value is 0.04\% away from the RPA result in Ref.~\cite{Johnson1986}. It is worth noting that the RPA value is only 1\% away from the more accurate CCSDvT result \cite{Porsev2009,Porsev2010}.

\section{Parity-mixed coupled-cluster method}\label{Sec:PMCC}
In the previous section, we demonstrated the utility of the PM-DHF basis sets in relativistic many-body calculations by computing the \Cs\ $6S_{1/2}-7S_{1/2}$ PNC transition amplitude in the all-order RPA method. The RPA result \eqref{RPA_res} includes all second-order MBPT corrections to matrix elements, but omits important third-order effects including the so-called Brueckner-orbital diagrams whose contributions are numerically as important as the RPA ones \cite{Lowdin1958,Blundell1992,Johnson1996}. The task of accounting for these higher-order MBPT corrections can be systematically carried out by means of the coupled-cluster (CC) method~\cite{CoeKum60,Ciz66}. For example, it is well-known that a CC formalism which includes singles, doubles and valence triples (CCSDvT) particle-hole excitations from the lowest-order state is complete through the fourth order of MBPT for energies and through the fifth order for matrix elements \cite{PorDer06Na,DerPorBel08}. 

The goal of this section is to outline a PM generalization to the PP-CCSDvT method used in Refs.~\cite{Porsev2009,Porsev2010}, where the conventional PP-DHF basis sets were employed. A labor-intensive numerical implementation of the method discussed here will be the subject of our future work. Since there are multiple implementations of relativistic PP-CC methods, especially in the quantum chemistry community, our theoretical formulation may be useful in the work of other groups as well.

There are several advantages to the PM-CC formulation. First of all, the PP-CC codes are already available,
and we outline the strategy for a relatively straightforward generalization of these codes. For example, the CCSDvT method reproduces the relevant atomic properties at a few 0.1\% accuracy level, therefore 
the PM-CCSDvT method (barring implementation errors) should at least be as accurate. Moreover, 
as mentioned in Sec.~\ref{Introduction}, since the lowest order PM-DHF result is only 3\% away  from the more accurate CCSDvT value~\cite{Porsev2009,Porsev2010}, the correlation corrections in the PM approach are substantially smaller than in CCSDvT  and, hence, a greater accuracy can be expected. In addition, the PM-CC formulation avoids directly summing over intermediate states in expressions for
parity non-conserving amplitudes, as in the original  PP-CCSDvT method. This reduces theoretical uncertainties associated with highly-excited and core-excited intermediate states, a subject of controversy \cite{Porsev2009,Porsev2010,Dzuba2012,SahDasSoi2021-CsPNC}. 

We begin our discussion by going back to the second-quantized form of the full electronic Hamiltonian $H'$, Eq.~\eqref{2quant},
\begin{align}\label{Eq:SecQuantH}
        H' &=  H'_0 + G' \nonumber\\
          &= \sum_i \varepsilon'_i N[a'^\dagger_i a'_i] + \frac12\sum_{ijkl} g'_{ijkl} N[a'^\dagger_i a'^\dagger_j a'_l a'_k ] \,,
\end{align}
where we have dropped the one-particle term $\sum_{ij}\left(V'_{\rm HF}-U'\right)_{ij}N[a'^\dagger_ia'_j]$ which vanishes due to our choice of the potential $U'=V'_{\rm HF}$. 

In the CC formalism, the exact many-body eigen-state $|\Psi'_v\rangle$ of the Hamiltonian $H'$ can be represented as 
\begin{align}\label{Eq:PsivOmega}
        |\Psi'_v\rangle &= \Omega' |\Psi_v^{(0)}\rangle = N[ \exp(K') ]\, |\Psi'^{(0)}_v\rangle\nonumber\\
                    &= \left( 1 + K' + \frac{1}{2!} N[K'^2] + \ldots \right)
\, |\Psi'^{(0)}_v\rangle \, ,
\end{align}
where $\Omega'$ is the wave operator, $|\Psi'^{(0)}_v \rangle$ is again the lowest-order PM-DHF state and the cluster operator $K'$ is expressed in terms of connected diagrams of the wave operator \cite{lindgren2012atomic}. In the CCSDvT approach, the cluster operator $K'$ is approximated by
\begin{align}\label{Eq:KCCSDvT}
         K' &=\sum_n(K'_c)_n+\sum_n(K'_v)_n\nonumber\\
         &\approx S'_c + D'_c + S'_v + D'_v +T'_v\nonumber \\
           &= \sum_{ma}\rho'_{ma}a'^\dagger_m a'_a + \frac{1}{2!}\sum_{mnab}\rho'_{mnab}a'^\dagger_ma'^\dagger_na'_ba'_a\nonumber\\
           &+\sum_{m\neq v}\rho'_{mv}a'^\dagger_m a'_v+ \frac{1}{2!}\sum_{mna}\rho'_{mnva}a'^\dagger_ma'^\dagger_na'_aa'_v\nonumber\\
           &+ \frac{1}{3!}\sum_{mnrab}\rho'_{mnrvab}a'^\dagger_ma'^\dagger_na'^\dagger_ra'_ba'_aa'_v\nonumber\\
           &=\begin{array}{l}
             \includegraphics*[scale=0.4]{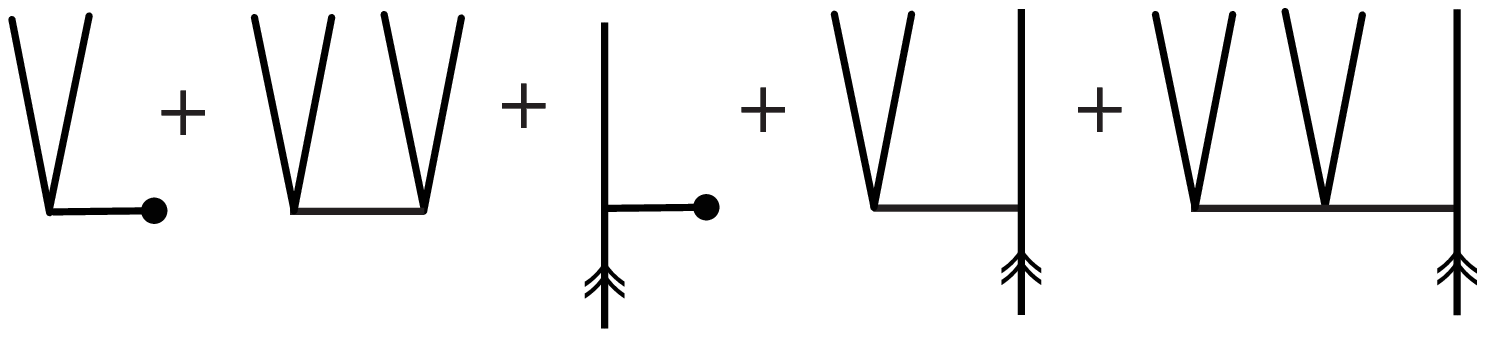}
             \end{array}\,,
    \end{align}
with the double-headed arrow representing the valence state.
Here $S'_v\equiv(K'_v)_1$ and $D'_v\equiv(K'_v)_2$ are the PM valence singles and doubles, $S'_c\equiv(K'_c)_1$ and $D'_c\equiv(K'_c)_2$ are the PM core singles and doubles and $T'_v\equiv(K'_v)_3$ is the PM valence triples. Note that here, they are expressed in terms of the PM creation and annihilation operators $a'^\dagger_i$ and $a'_i$, in contrast to the conventional CC approach where the cluster operators are expressed in terms of the PP creation and annihilation operators.

The goal of a CC calculation is to compute the cluster amplitudes $\rho'$ in Eq.~\eqref{Eq:KCCSDvT}. These amplitudes may be found from the Bloch equation \cite{lindgren2012atomic} specialized for univalent systems \cite{DerEmmons2002}
\begin{equation}\label{Bloch_eqn}
    \begin{aligned}
    \left(\varepsilon'_v-H'_0\right)(K_c)_n&=\left\{Q'G'\Omega'\right\}_{{\rm connected},n}\,,\\
    \left(\varepsilon'_v+\delta E_v-H'_0\right)(K_v)_n&=\left\{Q'G'\Omega'\right\}_{{\rm connected},n}\,,
    \end{aligned}
\end{equation}
where the valence correlation energy is given by
\begin{equation}\label{delta_Ev}
    \delta E_v= \bra{\Psi^{(0)}_v}G\Omega\ket{\Psi^{(0)}_v}\,,
\end{equation}
and $Q'\equiv1-\ket{\Psi'^{(0)}_v}\bra{\Psi'^{(0)}_v}$ is a projection operator onto the space spanned by the PM excited states. The subscript ``connected'' means that only connected diagrams are retained on the right-hand sides of Eqs.~\eqref{Bloch_eqn}.

It is worth stressing that we have used for the energy correction $\delta E_v$, Eq.~\eqref{delta_Ev} the formula in the conventional PP-CC scheme. This is justified since the effects of the weak interaction on energies are $O(\eta^2)$. Although this is intuitively clear, we shall provide a rigorous proof once we have presented the parity decomposition of the PM cluster amplitudes, c.f.~Eqs.~\eqref{corre_energies}.

Since the commutation and contraction relations among the PM operators $a'^\dagger_i$ and $a'_i$ are identical to those for the PP operators $a^\dagger_i$ and $a_i$, the structure of Eqs.~\eqref{Bloch_eqn} for the PM cluster amplitudes is the same as for the PP amplitudes.~\cite{PorDer06Na,PalSafJoh07,Porsev2009,Porsev2010}. In appendix \ref{CC_eqns}, we collect these equations and list them in their explicit form. 

Note that in presenting the PM-CC equations, we have used anti-symmetrized combinations for doubles \begin{subequations}
\begin{align}
    \tilde{\rho}'_{mnab}&\equiv\rho'_{mnab}-\rho'_{mnba}=\rho'_{mnab}-\rho'_{nmab}\nonumber\\
    &=\frac12\left(\rho'_{mnab}+\rho'_{nmba}-\rho'_{mnba}-\rho'_{nmab}\right)\,,\label{76a}\\
    \tilde{\rho}'_{mnva}&\equiv\rho'_{mnva}-\rho'_{nmva}\,,
\end{align}
\end{subequations}
which have the following symmetry properties
\begin{subequations}\label{doub_rho_symm}
\begin{align}
    \tilde{\rho}'_{mnab}&=-\tilde{\rho}'_{nmab}=-\tilde{\rho}'_{mnba}\,,\\
    \tilde{\rho}'_{mnva}&=-\tilde{\rho}'_{nmva}\,,
\end{align}
\end{subequations}
and the fully anti-symmetrized valence triples amplitude $\tilde{\rho}'_{mnrvab}$, which is anti-symmetric with respect to any permutation of the indices $mnr$ or $ab$, e.g.,
\begin{align}\label{trip_rho_symm}
    \tilde{\rho}'_{mnrvab}&=-\tilde{\rho}'_{nmrvab}=-\tilde{\rho}'_{mrnvab}\nonumber\\
    &=-\tilde{\rho}'_{mnrvba}=\tilde{\rho}'_{mrnvba}=\ldots\,.
\end{align}
The symmetry properties \eqref{doub_rho_symm} and \eqref{trip_rho_symm} are useful for simplifying the CC codes.

Let us now discuss the structure of the PM-CC amplitudes. A general knowledge of this structure will prove useful for the implementation of the PM-CC equations. We begin with the PM single amplitudes $\rho'_{ma}$ and $\rho'_{mv}$. 

In the conventional CC approach where a PP single-particle basis is used, the single amplitudes have the angular decomposition
\begin{equation}\label{CC-singles}
    \begin{aligned}
        \rho_{ma}=\delta_{\kappa_m\kappa_a}\delta_{m_mm_a}S(ma)\,,\\
        \rho_{mv}=\delta_{\kappa_m\kappa_v}\delta_{m_mm_v}S(mv)\,,\\
    \end{aligned}
\end{equation}
where $S(ma)$ and $S(mv)$ are PP reduced single amplitudes.
In the current formalism with PM single-particle orbitals, Eqs.~\eqref{CC-singles} may be generalized by appending to $\rho_{ma}$ and $\rho_{mv}$ $P$-odd imaginary components as
\begin{equation}\label{CC-singles-PM}
    \begin{aligned}
        \rho'_{ma}=\delta_{m_mm_a}\left[\delta_{\kappa_m\kappa_a}S(ma)+\im\eta\delta_{\kappa_m,-\kappa_a}S''(ma)\right]\,,\\
        \rho'_{mv}=\delta_{m_mm_v}\left[\delta_{\kappa_m\kappa_v}S(mv)+\im\eta\delta_{\kappa_m,-\kappa_v}S''(mv)\right]\,,\\
    \end{aligned}
\end{equation}
where $S''(ma)$ and $S''(mv)$ are PNC singles amplitudes. The fact that a PM singles amplitude indeed breaks down into two mutually exclusive components, a $P$-even real part and a $P$-odd imaginary part, is proved in appendix \ref{App_B}.

Next, we consider the PM double amplitudes $\rho'_{mnab}$ and $\rho'_{mnvb}$. As discussed in Sec.~\ref{Sec:MatElsPMBasis}, the PM Coulomb matrix element $\tilde{g}'_{ijkl}$ retains its angular structure, Eq.~\eqref{g_ijkl_PNC}, but the reduced matrix element $Z'_L(ijkl)$ acquires a $P$-odd imaginary part. Since the PP double excitation coefficients have the same angular decomposition as the Coulomb matrix elements and the weak interaction conserves total angular momentum, one may decompose the PM double amplitudes as 
\begin{equation}
\begin{aligned}
    \tilde{\rho}'_{mnab} &= \sum_LJ_L(mnab)\tilde{S}'_L(mnab)\,,\\
    \tilde{\rho}'_{mnvb} &= \sum_LJ_L(mnvb)\tilde{S}'_L(mnvb)\,,
\end{aligned}
\end{equation}
where $\tilde{S}'_L(mnab)$ and $\tilde{S}'_L(mnvb)$ are the reduced double amplitudes. 

Similarly to the case of the PM single amplitudes, it may be shown that the the reduced double amplitudes decompose into real and imaginary parts
\begin{equation}
\begin{aligned}
    \tilde{S}'_L(mnab)&=\tilde{S}_L(mnab)+\im\eta\tilde{S}''_L(mnab)\,,\\
    \tilde{S}'_L(mnvb)&=\tilde{S}_L(mnvb)+\im\eta\tilde{S}''_L(mnvb)\,.
\end{aligned}
\end{equation}
Here the real part $\tilde{S}_L(mnab)$ vanishes if $\ell_m+\ell_n+\ell_a+\ell_b$ is odd whereas the imaginary part $\im\eta\tilde{S}''_L(mnab)$ vanishes if $\ell_m+\ell_n+\ell_a+\ell_b$ is even. The same rules apply for $\tilde{S}_L(mnvb)$ and $\im\eta\tilde{S}''_L(mnvb)$.

The proof that the reduced double amplitudes indeed separate into mutually exclusive real and imaginary parts with opposite parity selection rules proceeds in a similar manner as for single-excitation coefficients, c.f.~appendix \ref{App_B}.

Finally, we consider the  PM valence triple amplitudes. Again, since the weak interaction does not break the total angular momentum selection rules, $\tilde{\rho}'_{mnrvab}$ has the following angular decomposition \cite{PorDer06Na}
\begin{equation}\label{rho_trip}
    \tilde{\rho}'_{mnrvab}=\sum_{LL'h}\begin{array}{l}
             \includegraphics*[scale=0.5]{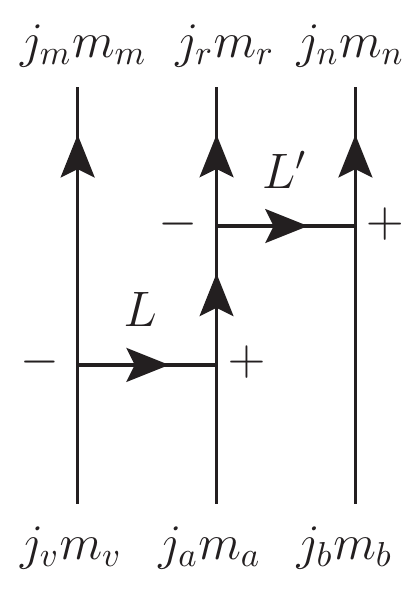}
             \end{array}\tilde{S}'_{LL'h}(mnrvab)\,,
\end{equation}
where $h$ is a half integer coupling angular momentum and $L$
and $L'$ are integer coupling momenta. The formula for writing the algebraic expression corresponding to the angular diagram in Eq.~\eqref{rho_trip} may be found in Ref.~\cite{lindgren2012atomic}.

The PM reduced triple amplitude $\tilde{S}'_{LL'h}(mnrvab)$ does not depend on the magnetic quantum numbers. Similar to the reduced double amplitudes, $\tilde{S}'_{LL'h}(mnrvab)$ may be decomposed into a $P$-even real part and a $P$-odd imaginary part, i.e.,
\begin{align}
    \tilde{S}'_{LL'h}(mnrvab) &= \tilde{S}_{LL'h}(mnrvab) \nonumber\\
    &+ \im\eta\tilde{S}''_{LL'h}(mnrvab)\,,
\end{align}
where $\tilde{S}_{LL'h}(mnrvab)$ vanishes if $\ell_m+\ell_n+\ell_r+\ell_v+\ell_a+\ell_b$ is odd and $\tilde{S}''_{LL'h}(mnrvab)$ vanishes if $\ell_m+\ell_n+\ell_r+\ell_v+\ell_a+\ell_b$ is even. The proof that valence triple amplitudes separate into mutually exclusive real and imaginary parts of opposite parities proceeds in a similar manner as for single- and double-excitation coefficients, c.f.~appendix \ref{App_B}.%, although the algebra is considerably lengthier. As a result, we do not present this proof in this paper either.

We have described the angular and parity structure of the PM-CC single, double, and valence triples amplitudes. Let us now turn our attention to the correlation corrections to the energy expressed in the PM basis. Up to the level of valence triples, the valence energy correction may be written as
\begin{equation}
\delta E'_v = \delta E'_\mathrm{SD} + \delta E'_\mathrm{CC} + \delta E'_\mathrm{vT} \,,
\label{Eq:delta_Ev}
\end{equation}
where $\delta E'_\mathrm{SD}$ represents the linear singles-doubles corrections
\begin{align}\label{dESD}
    \delta E'_\mathrm{SD}&=\sum_{ma}\tilde{g}'_{vavm}\rho'_{ma}\nonumber\\
    &+\frac12\sum_{mab}\tilde{g}'_{abvm}\tilde{\rho}'_{mvab}+\frac12\sum_{mnb}\tilde{g}'_{vbmn}\tilde{\rho}'_{mnvb}\,,
\end{align}
while $\delta E'_\mathrm{CC}$ contains contributions from nonlinear singles-doubles terms
\begin{align}
\delta E'_\mathrm{CC} &= \sum_{abnr}\tilde{g}'_{abnr}
\left[ \rho'_{vb} \rho'_{nrva} - \rho'_{nb} \tilde{\rho}'_{vrva} -
       \rho'_{nv} \rho'_{vrab} \right] \nonumber\\
&+\sum_{anr}\tilde{g}'_{avnr} \rho'_{na} \rho'_{rv} +
\sum_{abn}\tilde{g}'_{abnv} \rho'_{va} \rho'_{nb}\,, \label{dECC}
\end{align}
and $\delta E'_\mathrm{vT}$ is the valence triples term
\begin{align}
\delta E'_\mathrm{vT} &= \frac{1}{2}\sum_{abmn}g'_{abmn}
\tilde{\rho}'_{vmnvab}\,. \label{dEvT}
\end{align}

For completeness, we also present the correlation correction to the core energy $\delta E'_c$, although it is not needed in the CC calculations. Since we do not include core triples in our formalism, the correlation correction to the core energy has the form
\begin{equation}\label{core_ener_corr}
    \delta E'_c=\frac12\sum_{abmn}g'_{abmn}\tilde{\rho}_{mnab}+\frac12\sum_{abmn}\tilde{g}'_{abmn}\rho_{ma}\rho_{nb}\,.
\end{equation}

Physically, the energy corrections presented here must be real-valued. Nevertheless, that they are so is not immediately clear from Eqs.~\eqref{dESD},~\eqref{dECC},~\eqref{dEvT} and \eqref{core_ener_corr} alone. However, once the decomposition of the PM Coulomb matrix elements and CC amplitudes into real and imaginary parts of opposite parities is taken into account, the reality of the CC energy corrections becomes apparent. 

For example, consider the term $\sum_{ma}\tilde{g}_{vavm}\rho_{ma}$ in Eq.~\eqref{dESD}. Since $v$ appears twice in $\tilde{g}_{vavm}$, its (nominal) parity is $(-1)^{\ell_m+\ell_a}$, which is the same as that of $\rho_{ma}$. Thus, $\tilde{g}_{vavm}$ and $\rho_{ma}$ are either both real or imaginary simultaneously so their product is always real. We may also consider, for example, the term $\sum_{abnr}\tilde{g}_{abnr}\rho_{vb} \rho_{nrva}$ in Eq.~\eqref{dECC}. Suppose now that $\ell_a+\ell_n+\ell_r$ is odd and that $\ell_b$ and $\ell_v$ are even. Then $\tilde{g}_{abnr}$ and $\rho_{nrva}$ are both imaginary while $\rho_{vb}$ is real. As a result, the product of these three terms are real and the same argument applies to other cases. The upshot here is that if the total parities of several quantities is even then their product is real. Since the indices of the terms contributing to the correlation energy always appear in pairs, the total parity of each contribution is even and thus they are all real. 

Moreover, since each imaginary quantity comes with the small factor $\eta$, the PM correlation corrections to the energies of the core $E_c$ and a valence state $E_v$ are given by
\begin{equation}\label{corre_energies}
\begin{aligned}
    \delta E'_c = \delta E_c + O(\eta^2)\,,\\
    \delta E'_v = \delta E_v + O(\eta^2)\,,
\end{aligned}
\end{equation}
where $\delta E_c$ and $\delta E_v$ are the correlation corrections calculated using PP bases. This fact is in agreement with the general observation that the weak interaction does not produce energy shifts up to $O(\eta^2)$.

In a conventional PP-CC scheme, the correlation energies are used as a test for the convergence pattern of the CC amplitudes. Analogously, Eqs.~\eqref{corre_energies} show that in a PM-CC calculation, the correlation energies can be used to test the convergence of the real parts of the CC amplitudes. They do not, however, provide information about the convergence of the imaginary parts. We control the convergence patterns of these $P$-odd components by directly observing the largest change from iteration to iteration.

The fact that a cluster amplitude's real and imaginary parts are of opposite parities allows us to formulate a strategy for implementing the PM-CCSDvT code as follows. First, the conventional PP-CC program is executed until it converges. The resulting PP cluster amplitudes from this program are, up to $O(\eta^2)$, the real components of their PM counterparts. These PP amplitudes are then used as the initial values in a modified PM-CC code. This modified code uses the complex-valued PM matrix elements $z'_{ij}$ and $\tilde{g}'_{ijkl}$ (with parity selection rules modified accordingly) to compute the imaginary PNC part of the PM cluster amplitudes. The convergence of these imaginary parts are checked via their relative changes from iteration to iteration.

Once the PM cluster amplitudes and
correlation energies have been found, one may use the obtained wave functions for two valence states $\Psi'_w$ and $\Psi'_v$ to evaluate various matrix elements, such as that of the electric dipole operator entering PNC amplitude,
\begin{equation}
 Z'_{wv} =
 \frac{ \langle \Psi'_w | \sum_{ij} \langle i' | z | j'\rangle \, a'^\dagger_i a'_j | \Psi'_v \rangle }
 { \sqrt{ \langle \Psi'_w | \Psi'_w \rangle \langle \Psi'_v | \Psi'_v \rangle  }}
 \label{Eq:Zmel} \, .
\end{equation}
The corresponding CCSDvT expressions are given in Ref.~\cite{PorDer06Na}. The ``dressing'' of lines and vertices in expressions for matrix elements is the same as discussed in Ref.~\cite{DerPor05}. The only difference is that all the PP quantities are to be replaced by PM ones. 

\section{Discussion}\label{Conclusions}
We have discussed how a conventional coupled-cluster (CC) calculation which uses parity-proper (PP) single-electron basis functions may be generalized to use parity-mixed (PM) basis functions instead. In this PM version of the CC method, the parity non-conserving (PNC) electron-nucleus weak interaction is incorporated into the zeroth-order single-electron Dirac-Hartree-Fock (DHF) Hamiltonian. Such a PM-CC formulation has the advantage over the traditional PP-CC method for several reasons. 

Firstly, in a conventional PP-CC calculation of the PNC amplitude where the sum-over-states approach is used, c.f.~Eq.~\eqref{SoS_Eqn}, contributions to the PNC amplitude are often split into a ``main'' term, coming from low-lying excited states, and a ``tail'' term, coming from highly-excited and core-excited intermediate states. A typical breakdown of these contributions \cite{Porsev2009,Porsev2010} is given in Table \ref{SoS_table}. 

It may be observed from Table \ref{SoS_table} that although the ``main'' and ``tail'' terms in the CCSDvT method have the same absolute uncertainty, the fractional inaccuracy of the former is at the level of 0.2\%, whereas that of the latter reaches 10\%. In the PM-CC approach, summing over states and thus the artificial separation into ``main'' and ``tail'' terms are avoided. As a result, the fractional inaccuracies of all the contributions are anticipated to be at the level of $0.2\%$. Since the ``tail'' contributes only 2\% to the PV amplitude, this means that one of the largest sources of error in Table \ref{SoS_table} will be effectively removed. Thereby, with the PM-CCSDvT approach, we anticipate improving the current 0.5\% theoretical uncertainty \cite{Dzuba2012} to $\sim$ 0.2\%, reaching the new improved accuracy level in the low-energy test of the electroweak sector of the Standard Model.

\begin{table}[htb]
\begin{tabular}{ll}
\hline\hline
\multicolumn{2}{c}{Coulomb interaction corrections}                            \\
\hline
Main ($n=6 - 9$)               & 0.8823(18)                        \\
{\color{red} Tail}    & {\color{red} 0.0175(18)} \\
Total correlated               & 0.8998(25)                        \\
\hline
\multicolumn{2}{c}{Other corrections}                                    \\
\hline
Breit, Ref.~\cite{Derevianko2000}                   & -0.0054(5)                        \\
QED, Ref.~\cite{Shabaev2005}                      & -0.0024(3)                        \\
Neutron skin, Ref.~\cite{derevianko_2001_Breit_NeutronSkin}            & -0.0017(5)                        \\
$e - e$ weak interaction, Ref.~\cite{Blundell1990} & 0.0003                            \\
\hline
Final                          & 0.8906(26)                       \\
\hline\hline
\end{tabular}
\caption{Contributions to the parity violating amplitude EPV for the $6S_{1/2}\rightarrow 7S_{1/2}$ transition in ${}^{133}$Cs in units of $10^{-11}i|e|a_0Q_W/N$. Here, $N = 78$ is the number of neutrons in the ${}^{133}$Cs nucleus. The results are from the CCSDvT calculations \cite{Porsev2009,Porsev2010} in the sum-over-state approach. Improving the accuracy of the tail contribution (shown in red) is the goal of the current work.}
\label{SoS_table}
\end{table}

Secondly, the lowest-order DHF result in this PM approach is only 3\% away from the more accurate CCSDvT value. This is to be compared with the traditional DHF result which is off by 18\%.
% correlated CCSDvT 0.8998(24); PM-DHF 0.92701; DHF 0.73946 
This indicates that the correlation corrections in the PM approach are substantially smaller than in the conventional PP method. Depending on the MBPT convergence pattern, one can generically expect an  improved theoretical accuracy. In addition, the upgrade of existing and well-tested large-scale PP-CC codes to PM-CC ones is relatively straightforward.

The implementation of a PM-CC code requires a basis of PM single-electron orbitals, which are eigen-states of the PM-DHF Hamiltonian~\eqref{Hamiltonian'}. In this paper, we presented several methods through which these eigen-states may be obtained with high accuracy. We note here that the symmetry of Eq.~\eqref{SoS_Eqn} with respect to the exchange $D_z\leftrightarrow H_W$ suggests an alternative approach to the APV problem: instead of using the Hamiltonian ~\eqref{Hamiltonian'}, one adds an operator $\lambda D_z$ ($\lambda$ may be thought of as the strength of an external electric field ${\bf E}=\lambda\hat{\bf z}$) to the PP atomic Hamiltonian $H$ and solves for the eigen-states $\Psi(\lambda)$ of $H+\lambda D_z$. With $\Psi(\lambda)$ obtained, one may then proceed to computing the expectation value $\bra{\Psi(\lambda)}H_W\ket{\Psi(\lambda)}$, whence the PNC transition amplitude \eqref{SoS_Eqn} may be calculated by taking the first derivative with respect to $\lambda$. This method has certain advantages such as the relatively simple form of the operator $D_z$. However, since $D_z$ is a tensor operator of rank one, as opposed to $H_W$ which is a pseudoscalar, it can couple orbitals with different total angular momenta, thus leading to a drastic increase in the number of allowed angular channels in a CC calculation.%We demonstrated that the lowest order value of the PNC transition amplitude calculated using the PM basis orbitals is significantly closer to the ``nominal'' value \cite{Porsev2009,Porsev2010} than the conventional PP-DHF results, which indicates that correlation effects in the PM basis is smaller and may thus be accounted for more accurately. 

With the PM bases obtained, we proceeded to computing the PM matrix elements of inter-electron Coulomb interaction and the electric dipole operator. The former are needed for the computation of correlation corrections to the single-electron wave functions while the latter are needed to calculate the PNC amplitude. We demonstrated the numerical accuracy of our PM approach by using these PM matrix elements in a random-phase approximation (RPA) calculation of the PNC amplitude, obtaining a 0.04\% agreement with a previous RPA result \cite{Johnson1985}.

Finally, we presented the extension of the conventional PP-CC method to a PM-CC formalism.  We also proved rigorously that a PM cluster amplitude is a complex number which decomposes into mutually exclusive $P$-even real part and $P$-odd imaginary part. An immediate consequence of this decomposition is that the correlations energies computed from these amplitudes are, reassuringly, real. More importantly, that a PM cluster amplitude is either real or imaginary depending on its nominal parity allows us to formulate a strategy for the PM-CC program.

%We first iterate the conventional PP-CC to obtain the PP cluster amplitudes and correlation energy corrections, which, up to $O(\eta^2)$ are, respectively, the real parts of the PM cluster amplitudes and the PM correlation energy corrections. These will then be used as the starting point for our PM-CC iteration to obtain the PNC imaginary parts of the PM cluster amplitudes. Once the PM cluster amplitudes are found, one may proceed to finding the many-body wave functions for the two valence states $\Psi'_w$ and $\Psi'_v$ then to evaluating various quantities of interest, such as that of the PNC amplitude, Eq.~\eqref{Eq:Zmel}.

A full implementation of the PM-CCSDvT calculation based on the strategy mapped out here will be a subject of our future work. The result of this computation will help with the interpretation of the next generation searches for new physics with atomic parity violation (APV) \cite{Antypas2013,choi2018gain}. In addition, since there are multiple implementations of relativistic PP-CC methods, especially in the quantum chemistry community, our theoretical formulation may be useful in the work of other groups.

\section*{Acknowledgements}
We would like to thank Walter Johnson for valuable discussion.
This work was supported in part by the U.S. National Science Foundation grant PHY-1912465, by the Sara Louise Hartman endowed professorship in Physics, and by the Center for Fundamental Physics at Northwestern University.

\appendix
\section{Parity-mixed coupled-cluster equations}\label{CC_eqns}
In this appendix, we present the equations for the singles, doubles and valence triples CC amplitudes in their explicit form. For ease of presentation, we shall suppress all primes on the quantities involved. It should still be understood, however, that the quantities appearing here are of PM character and are generally complex numbers. For convenient, we will use the notation $\varepsilon_{ijk\dots}\equiv\varepsilon_i+\varepsilon_j+\varepsilon_k+\dots$ to denote sums of single-electron energies \cite{PorDer06Na,PalSafJoh07}.

The equation for the core single-excitation coefficients reads
\begin{equation}\label{core_single_eqn}
\begin{aligned}
    (\varepsilon_a-\varepsilon_m)\rho_{ma} = X^s_{\rm SD} + \sum_{i=1}^3A^s_i\,,
\end{aligned}
\end{equation}
where $X^s_{\rm SD}$ is the linearized singles-doubles term
\begin{align}\label{XSDs}
    X^s_{\rm SD} &= \sum_{bn}\tilde{g}_{mban}\rho_{nb}+\frac12\sum_{bnr}\tilde{g}_{mbnr}\tilde{\rho}_{nrab}\nonumber\\
    &-\frac12\sum_{bcn}\tilde{g}_{bcan}\tilde{\rho}_{mnbc}\,,
\end{align}
while $A^s_{1,2,3}$ are all the nonlinear singles-doubles terms
\begin{align}
    A^s_1 &= \sum_{drs}\tilde{g}_{mdrs}\rho_{ra}\rho_{sd}-\sum_{cds}\tilde{g}_{cdas}\rho_{mc}\rho_{sd}\,,
\end{align}
\begin{align}
    A^s_2 &= - \frac12\sum_{cdrs}\tilde{g}_{cdrs}\tilde{\rho}_{rsda}\rho_{mc} - \frac12\sum_{cdrs}\tilde{g}_{cdsr}\tilde{\rho}_{smcd}\rho_{ra}\nonumber\\
    &+\sum_{cdrs}\tilde{g}_{cdrs}\tilde{\rho}_{rmca}\rho_{sd}\,,
\end{align}
\begin{align}
    A^s_3&=-\sum_{cdrs}\tilde{g}_{cdsr}\rho_{mc}\rho_{rd}\rho_{sa}\,.
\end{align}

The equation for the core double-excitation coefficients reads
\begin{equation}\label{core_double_eqn}
    (\varepsilon_{ab}-\varepsilon_{mn})\rho_{mnab}=X^d_{\rm SD} + \sum_{i=1}^6A^d_i\,,
\end{equation}
where the linearized singles-doubles term is given by
\begin{align}\label{XdSD}
    X^d_{\rm SD} &= g_{mnab} +\frac14 \sum_{cd}\tilde{g}_{cdab}\tilde{\rho}_{mncd}+\frac14\sum_{rs}\tilde{g}_{mnrs}\tilde{\rho}_{rsab}\nonumber\\
    &+ \left[\sum_rg_{mnrb}\rho_{ra}-\sum_cg_{cnab}\rho_{mc}\right.\nonumber\\
    &\left.+\sum_{cr}\tilde{g}_{cnrb}\tilde{\rho}_{mrac}+ \begin{pmatrix}a\leftrightarrow b\\m\leftrightarrow n\end{pmatrix}\right]\,,
\end{align}
and the nonlinear singles-doubles terms are given by
\begin{align}\label{Ad1}
    A^d_1&=\sum_{rs}g_{mnrs}\rho_{ra}\rho_{sb}+\sum_{cd}g_{cdab}\rho_{mc}\rho_{nd}\nonumber\\
    &-\left[\sum_{dr}\tilde{g}_{mdar}\rho_{rb}\rho_{nd}+\begin{pmatrix}a\leftrightarrow b\\m\leftrightarrow n\end{pmatrix}\right]\,,
\end{align}
\begin{align}\label{Ad2}
    A^d_2&=\left[\sum_{cdr}\tilde{g}_{cdrb}\rho_{nd}\tilde{\rho}_{rmca} -\sum_{cdr}\tilde{g}_{cdar}\rho_{rd}\rho_{mncb} \right.\nonumber\\
    &+\sum_{cdr}g_{cdra}\rho_{rb}\rho_{nmcd} + \sum_{crs}\tilde{g}_{ncrs}\rho_{rb}\tilde{\rho}_{smca}\nonumber\\
    &+\sum_{crs}\tilde{g}_{ncrs}\rho_{sc}\rho_{mrab}-\frac14\sum_{crs}\tilde{g}_{ncrs}\rho_{mc}\tilde{\rho}_{srab}\nonumber\\&\left.+\begin{pmatrix}a\leftrightarrow b\\m\leftrightarrow n\end{pmatrix}\right]\,,
\end{align}
\begin{align}
    A^d_3 &= \left[\sum_{cdr}g_{cdar}\rho_{nd}\rho_{mc}\rho_{rb}-\sum_{crs}g_{mcrs}\rho_{nc}\rho_{ra}\rho_{sb}\right.\nonumber\\
    &\left.+\begin{pmatrix}a\leftrightarrow b\\m\leftrightarrow n\end{pmatrix}\right]\,,
\end{align}
\begin{align}
    A_{4}^{d}=& \sum_{cdtu} g_{cdtu} \rho_{tuab} \rho_{mncd}+\sum_{cdtu} \tilde{g}_{cdtu} \tilde{\rho}_{mtac} \tilde{\rho}_{undb} \nonumber\\
    &-\left[\sum_{cdtu} \tilde{g}_{cdtu}\left(\rho_{tubd} \rho_{mnac}+\rho_{mucd} \rho_{ntba}\right)\right.\nonumber\\
    &\left.+\begin{pmatrix}a\leftrightarrow b\\m\leftrightarrow n\end{pmatrix}\right]\,,
\end{align}
\begin{align}
    A^d_5&=\sum_{cdtu}g_{cdtu}\left(\rho_{ta}\rho_{ub}\rho_{mncd} + \rho_{mc}\rho_{nd}\rho_{tuab} \right)\nonumber\\
    &-\left[\sum_{cdtu}\tilde{g}_{cdut}\rho_{tb}\rho_{uc}\rho_{mnad}+\sum_{cdtu}\tilde{g}_{cdtu}\rho_{tc}\rho_{nd}\rho_{muab}\right.\nonumber\\
    &\left.\sum_{cdtu}\tilde{g}_{cdtu}\rho_{tb}\rho_{nc}\tilde{\rho}_{muad}+\begin{pmatrix}a\leftrightarrow b\\m\leftrightarrow n\end{pmatrix}\right]\,,
\end{align}
\begin{align}\label{Ad6}
    A^d_6&=\sum_{cdtu}g_{cdtu}\rho_{ta}\rho_{ub}\rho_{mc}\rho_{nd}\,.
\end{align}

The equation for valence singles reads
\begin{align}
    (\varepsilon_v-\varepsilon_m+\delta E_v)\rho_{mv} &= \left(X^s_{\rm SD}\right)_{a\rightarrow v} \nonumber\\
    &+ \sum_{i=1}^3\left(A^s_i\right)_{a\rightarrow v} + B^s\,,
\end{align}
where $B^s$ stands for the contribution from valence triples
\begin{equation}
    B^s=\frac12\sum_{abnr}g_{abnr}\tilde{\rho}_{mnrvab}\,.
\end{equation}

The equation for valence doubles reads
\begin{align}
    (\varepsilon_{av}-\varepsilon_{mn}+\delta E_v)\rho_{mnva}&= \left(X^d_{\rm SD}\right)_{a\rightarrow v}\nonumber \\
    & + \sum_{i=1}^5\left(A^d_i\right)_{a\rightarrow v} + B^d\,,
\end{align}
where $B^d$ represents the effect of valence triples on valence doubles
\begin{align}
    B^d&= -\frac12\sum_{rbc}\left(
    g_{bcar} \tilde{\rho}_{mnrvbc}+ g_{bcvr} \tilde{\rho}_{nmrabc} \right)\nonumber\\
   &+ \frac12\sum_{rsb}\left(
    g_{bnrs} \tilde{\rho}_{msrvab}+
    g_{bmrs} \tilde{\rho}_{snrvab} \right) \,.
\end{align}

The equation for valence triples reads
\begin{equation}
(\varepsilon_{abv}-\varepsilon_{mnr}+\delta E_v)\tilde{\rho}_{mnrvab} = B^t_1 + B^t_2\,,
\end{equation}
where 
\begin{align}
B^t_1 & = +  \sum_s\left(
\tilde{g}_{nrsv}\tilde{\rho}_{msab}+\tilde{g}_{rmsv}%
\tilde{\rho}_{nsab}+\tilde{g}_{mnsv}\tilde{\rho}_{rsab}%
\right) \nonumber\\
&- \sum_c\left(
\tilde{g}_{mcva}\tilde{\rho}_{nrcb}-\tilde{g}_{mcvb}%
\tilde{\rho}_{nrca}+\tilde{g}_{ncva}\tilde{\rho}_{rmcb}\right.\nonumber\\
&\left.-\tilde{g}_{ncvb}\tilde{\rho}_{rmca}+\tilde{g}_{rcva}%
\tilde{\rho}_{mncb}-\tilde{g}_{rcvb}\tilde{\rho}_{mnca}
\right) \,,
\end{align}
\begin{align}
B^t_2 & =   \sum_c\left(
\tilde{g}_{mcab}\tilde{\rho}_{nrvc}+\tilde{g}_{ncab}%
\tilde{\rho}_{rmvc}+\tilde{g}_{rcab}\tilde{\rho}_{mnvc} \right) \nonumber \\
& +\sum_s\left(
\tilde{g}_{nrsb}\tilde{\rho}_{msva}-\tilde{g}_{nrsa}%
\tilde{\rho}_{msvb}+\tilde{g}_{rmsb}\tilde{\rho}_{nsva}\right.\nonumber\\
&\left.-\tilde{g}_{rmsa}\tilde{\rho}_{nsvb}+\tilde{g}_{mnsb}%
\tilde{\rho}_{rsva}-\tilde{g}_{mnsa}\tilde{\rho}_{rsvb} \right) \,.
\label{Eq:rho_mnrvab}
\end{align}

The formulas for the valence energy correction $\delta E_v$ in these equations were given in Eqs.~\eqref{dESD},~\eqref{dECC}, and \eqref{dEvT} in the main text.

\section{Parity decomposition of coupled-cluster amplitudes}\label{App_B}
In this appendix, we prove that PM cluster amplitudes decompose into real parts with even parities and imaginary parts with odd parities. More specifically, we show that (i) the singles amplitude $\rho'_{ma}$ ($\rho'_{mv}$) is purely real if the sum $\ell_m+\ell_a$ ($\ell_m+\ell_v$) is even but is purely imaginary if the sum is odd, (ii) the doubles amplitude $\rho'_{mnab}$ ($\rho'_{mnvb}$) is purely real if the sum $\ell_m+\ell_n+\ell_a+\ell_b$ ($\ell_m+\ell_n+\ell_a+\ell_b$) is even but is purely imaginary if the sum is odd, and (iii) the triples amplitude $\rho'_{mnrvab}$ is purely real if the sum $\ell_m+\ell_n+\ell_r+\ell_v+\ell_a+\ell_b$ is even but is purely imaginary if the sum is odd. Note that although we limit the current discussion to valence triples, the proof here applies to cluster amplitudes of all ranks.

There are several ways through which the selections rules imposed on the PM cluster amplitudes may be demonstrated. Here, present two such methods, namely, a proof by induction on the cluster equations (see appendix \ref{CC_eqns}) and a proof which uses the parity operator (see below).

\subsection{Proof using the cluster equations}
\subsubsection{Parity-proper cluster amplitudes}
We begin by deriving the parity selection rule imposed on the PP amplitudes. This serves as a starting point which motivates and generalizes well to the case of PM amplitudes. For definiteness, we concentrate on the PP core single and double amplitudes and prove by induction that 
\begin{subequations}\label{B1}
    \begin{align}
        \rho_{ma}&\propto{\rm mod}_2(\ell_m+\ell_a+1)\,,\label{B1a}\\
        \rho_{mnab}&\propto{\rm mod}_2(\ell_m+\ell_n+\ell_a+\ell_b+1)\,.\label{B1b}
    \end{align}
\end{subequations}
From the definition \eqref{76a}, one observes that $\tilde{\rho}_{mnab}$ has the same selection rule as $\rho_{mnab}$, Eq.~\eqref{B1b}. The selection rules for PP valence single, double, and triple amplitudes follow from those for core singles and doubles in a trivial way.

To prove the selection rules \eqref{B1} by induction, we consider solving the PP version of Eqs.~\eqref{core_single_eqn} and \eqref{core_double_eqn} iteratively. As initial values, we take $\rho^{(0)}_{ma}=0$ and $\rho^{(0)}_{mnab}=0$. Equations~\eqref{core_single_eqn} and \eqref{core_double_eqn} then give, after the first iteration
\begin{equation}
    \begin{aligned}
        \rho^{(1)}_{ma} &= 0\,,\\
        \rho^{(1)}_{mnab} &= g_{mnab}\,.
    \end{aligned}
\end{equation}
Equation \eqref{B1a} is satisfied trivially while Eq.~\eqref{B1b} is satisfied due to the selection rules on the Coulomb matrix elements $g_{mnab}$.

We now assume that Eqs.~\eqref{B1} are satisfied by $\rho^{(n)}_{ma}$ and $\rho^{(n)}_{mnab}$ for $n\geq 2$. Let us investigate the driving terms $(X_{\rm SD}^s)^{(n)}$ and $(A_i^s)^{(n)}$ on the right-hand side of the single equation, Eq.~\eqref{core_single_eqn}. A close inspection of these terms shows that they vanish if $\ell_a+\ell_m$ is odd. For example, the term $\tilde{g}^{(n)}_{mban}\rho^{(n)}_{nb}$ in Eq.~\eqref{XSDs} is nonzero only if both $
\ell_m+\ell_b+\ell_a+\ell_n$ and $\ell_n+\ell_b$ are even, which implies that $\ell_a+\ell_m$ is even. As a result, Eq.~\eqref{core_single_eqn} implies that
\begin{equation}
    \rho_{ma}^{(n+1)}=\frac{(X_{\rm SD}^s)^{(n)}+\sum_{i=1}^3(A_i^s)^{(n)}}{\epsilon_a-\epsilon_m}
\end{equation}
is zero if $\ell_a+\ell_m$ is odd.

Analogously, one may show that the driving terms $(X_{\rm SD}^d)^{(n)}$ and $(A_i^d)^{(n)}$ on the right-hand side of the double equation, Eq.~\eqref{core_double_eqn} vanish if $\ell_m+\ell_n+\ell_a+\ell_b$ is odd. For example, the term $\tilde{g}^{(n)}_{cdab}\tilde{\rho}^{(n)}_{mncd}$ in Eq.~\eqref{XdSD} vanishes unless $\ell_c+\ell_d+\ell_a+\ell_b$ and $\ell_m+\ell_n+\ell_c+\ell_d$ are both even. That these sums are both even is equivalent to requiring $\ell_m+\ell_n+\ell_a+\ell_b$ to be even. Similar arguments apply to all other terms in Eqs.~\eqref{XdSD}--\eqref{Ad6}. As a result, Eq.~\eqref{core_double_eqn} implies that
\begin{equation}\label{B4}
    \rho_{mnab}^{(n+1)}=\frac{(X_{\rm SD}^d)^{(n)}+\sum_{i=1}^6(A_i^d)^{(n)}}{\epsilon_{ab}-\epsilon_{mn}}
\end{equation}
is zero if $\ell_m+\ell_n+\ell_a+\ell_b$ is odd. Due to the definition \eqref{76a}, this selection rule for $\rho_{mnab}^{(n+1)}$ also applies to $\tilde{\rho}_{mnab}^{(n+1)}$.

By the principle of induction, we conclude that the selection rules \eqref{B1} are satisfied by all core single and double amplitudes. An argument along the same line shows that the valence singles, doubles and triples satisfy similar selection rules
\begin{equation}
	\begin{aligned}\label{B5}
		\rho_{mv}&\propto{\rm mod}_2(\ell_m+\ell_v+1)\,,\\
		\rho_{mnva}&\propto{\rm mod}_2(\ell_m+\ell_n+\ell_v+\ell_a+1)\,,\\
		\rho_{mnrvab}&\propto{\rm mod}_2(\ell_m+\ell_n+\ell_r+\ell_v+\ell_a+\ell_b+1)\,.
	\end{aligned}
\end{equation}
\subsubsection{Parity-mixed cluster amplitudes}
We have proved by induction the selection rules imposed on the PP cluster amplitudes from the PP cluster equations. Here, we generalize this method to show that the PM cluster amplitudes may be decomposed into real and imaginary parts with opposite parities. Again, we concentrate on the PM core singles and doubles, proving that 
\begin{subequations}\label{B6}
    \begin{align}
        {\rm Re}(\rho'_{ma})&\propto{\rm mod}_2(\ell_m+\ell_a+1)\,,\label{B6a}\\
        {\rm Im}(\rho'_{ma})&\propto{\rm mod}_2(\ell_m+\ell_a)\,,\label{B6b}\\
        {\rm Re}(\rho_{mnab})&\propto{\rm mod}_2(\ell_m+\ell_n+\ell_a+\ell_b+1)\,,\label{B6c}\\
        {\rm Im}(\rho_{mnab})&\propto{\rm mod}_2(\ell_m+\ell_n+\ell_a+\ell_b)\,.\label{B6d}
    \end{align}
\end{subequations}
Again, the selection rules for $\tilde{g}'_{mnab}$ are the same as those for $g'_{mnab}$ and the selection rules for PM valence single, double, and triple amplitudes follow directly from those for PM core singles and doubles.

Similarly to the PP case, we consider solving the PM version of Eqs.~\eqref{core_single_eqn} and \eqref{core_double_eqn} iteratively. As before, we take $\rho'^{(0)}_{ma}=0$ and $\rho'^{(0)}_{mnab}=0$. Equations~\eqref{core_single_eqn} and \eqref{core_double_eqn} then give, after the first iteration
\begin{equation}
    \begin{aligned}
        \rho'^{(1)}_{ma} &= 0\,,\\
        \rho'^{(1)}_{mnab} &= g'_{mnab}\,.
    \end{aligned}
\end{equation}
Equation \eqref{B6a} and \eqref{B6b} are satisfied trivially while Eqs.~\eqref{B6c} and~\eqref{B6d} are satisfied due to the selection rules on the PM Coulomb matrix elements $g'_{mnab}$ (see Sec.~\ref{Sec:MatElsPMBasis}). 

We now assume that Eqs.~\eqref{B6} are satisfied for $\rho'^{(n)}_{ma}$ and $\rho'^{(n)}_{mnab}$ with $n\geq 2$. It is then straightforward to show that the driving terms $(X'^s_{\rm SD})^{(n)}$ and $(A'^s_i)^{(n)}$ (we have used the prime to emphasize that these quantities are of PM character) on the right-hand side of the single equation, Eq.~\eqref{core_single_eqn}, are real if $\ell_a+\ell_m$ is even and are purely imaginary if $\ell_a+\ell_m$ is odd. For example, consider again the term $\tilde{g}'^{(n)}_{mban}\rho'^{(n)}_{nb}$ in Eq.~\eqref{XSDs}. By the induction assumptions, $\tilde{g}'^{(n)}_{mban}$ is real if $
\ell_m+\ell_b+\ell_a+\ell_n$ is even and is imaginary if $
\ell_m+\ell_b+\ell_a+\ell_n$ is odd. Similarly, $\rho'^{(n)}_{nb}$ is real if $
\ell_m+\ell_b+\ell_a+\ell_n$ is even and is imaginary if $
\ell_m+\ell_b+\ell_a+\ell_n$ is odd. As a result, the product $\tilde{g}'^{(n)}_{mban}\rho'^{(n)}_{nb}$ is real if $
\ell_m+\ell_b+\ell_a+\ell_n$ and $\ell_n+\ell_b$ have the same parity which can only be satisfied if $\ell_m+\ell_a$ is even. On the other hand, if $\ell_m+\ell_a$ is odd then $
\ell_m+\ell_b+\ell_a+\ell_n$ and $\ell_n+\ell_b$ have opposite parities and $\tilde{g}'^{(n)}_{mban}\rho'^{(n)}_{nb}$ is imaginary.

As a result, Eq.~\eqref{core_single_eqn} implies that
\begin{equation}
    \rho'^{(n+1)}_{ma}=\frac{(X'^s_{\rm SD})^{(n)}+\sum_{i=1}^3(A'^s_i)^{(n)}}{\epsilon_a-\epsilon_m}
\end{equation}
is purely real if $\ell_a+\ell_m$ is even and purely imaginary if $\ell_a+\ell_m$ is odd.

Analogously, it may be shown that the driving terms $(X'^d_{\rm SD})^{(n)}$ and $(A'^d_i)^{(n)}$ on the right-hand side of the double equation, Eq.~\eqref{core_double_eqn}, are real if $\ell_m+\ell_n+\ell_a+\ell_b$ is even and purely imaginary if this sum is odd. Consider again, for example, the term $\tilde{g}'^{(n)}_{cdab}\tilde{\rho}'^{(n)}_{mncd}$ in Eq.~\eqref{XdSD}. Whether this product is purely real or imaginary depends on the parities of its factors. This dependence is shown explicitly in Table~\ref{Tab_real_imag}. We observe from this table that $\tilde{g}'^{(n)}_{cdab}\tilde{\rho}'^{(n)}_{mncd}$ is purely real if $\ell_m+\ell_n+\ell_a+\ell_b$ is even and purely imaginary if $\ell_m+\ell_n+\ell_a+\ell_b$ is even is odd. Similar arguments apply to all other terms in Eqs.~\eqref{XdSD}--\eqref{Ad6}. As a result, one finds from Eq.~\eqref{core_double_eqn} that 
\begin{equation}
    \rho'^{(n+1)}_{mnab}=\frac{(X'^d_{\rm SD})^{(n)}+\sum_{i=1}^6(A'^d_i)^{(n)}}{\epsilon_{ab}-\epsilon_{mn}}
\end{equation}
is purely real if $\ell_m+\ell_n+\ell_a+\ell_b$ is even and purely imaginary if $\ell_m+\ell_n+\ell_a+\ell_b$ is even is odd. Due to the definition \eqref{76a}, the same selection rules hold for $\tilde{\rho}'^{(n+1)}_{mnab}$.

By the principle of induction, we conclude that the selection rules \eqref{B6} are satisfied by all PM core single and double amplitudes. An argument along the same line shows that the PM valence singles, doubles and triples satisfy similar conditions, i.e., 
\begin{equation}
    \begin{aligned}
        {\rm Re}(\rho_{mv})&\propto{\rm mod}_2(\ell_m+\ell_v+1)\,,\\
        {\rm Im}(\rho_{mv})&\propto{\rm mod}_2(\ell_m+\ell_v)\,,\\
        {\rm Re}(\rho_{mnva})&\propto{\rm mod}_2(\ell_m+\ell_n+\ell_v+\ell_a+1)\,,\\
        {\rm Im}(\rho_{mnva})&\propto{\rm mod}_2(\ell_m+\ell_n+\ell_v+\ell_a)\,,\\
        {\rm Re}(\rho_{mnrvab})&\propto{\rm mod}_2(\ell_m+\ell_n+\ell_r+\ell_v+\ell_a+\ell_b+1)\,,\\
        {\rm Im}(\rho_{mnrvab})&\propto{\rm mod}_2(\ell_m+\ell_n+\ell_r+\ell_v+\ell_a+\ell_b)\,,\\
    \end{aligned}
\end{equation}

\begin{table}[htb]
\begin{tabular}{ccccccc}
\hline\hline
\multirow{2}{*}{$\ell_a+\ell_b$} & \multirow{2}{*}{$\ell_c+\ell_d$} & \multirow{2}{*}{$\ell_m+\ell_n$} & \multirow{2}{*}{$\tilde{g}'^{(n)}_{cdab}$} & \multirow{2}{*}{$\tilde{\rho}'^{(n)}_{mncd}$} & $\ell_m+\ell_n$ & $\tilde{g}'^{(n)}_{cdab}$ \\
 & & & & & $+\ell_a+\ell_b$ & $\times\tilde{\rho}'^{(n)}_{mncd}$\\
\hline
e & e & e & r & r & e & r \\ \hline
e & e & o & r & i & o & i \\ \hline
e & o & e & i & i & e & r \\ \hline
e & o & o & i & r & o & i \\ \hline
o & e & e & i & r & o & i \\ \hline
o & e & o & i & i & e & r \\ \hline
o & o & e & r & i & o & i \\ \hline
o & o & o & r & r & e & r \\ \hline\hline
\end{tabular}
\caption{The dependence of the reality of the term $\tilde{g}'^{(n)}_{cdab}\tilde{\rho}'^{(n)}_{mncd}$ in Eq.~\eqref{XdSD} on the parity of the sum $\ell_m+\ell_n+\ell_a+\ell_b$. Here, ``e'' stands for even, ``o'' for odd, ``r'' for real, and ``i'' for imaginary.}\label{Tab_real_imag}
\end{table}

\subsection{Proof using the parity operator}
In the last subsection, we have proved that the PM cluster amplitudes decompose into real and imaginary parts of opposite parities by using induction. Here, we provide a different proof using the parity operator in second-quantized form (see Sec.~\ref{Sec:ParOpSecondQuantization})
\begin{align}\label{Eq:ParOp_2ndQuan}
   \Pi&= \sum_{\mu=0}^{N_e}\frac{1}{(\mu!)^2}\sum_{\{a\}\{m\}}(-1)^{\sum_{i=1}^\mu\ell_{a_i}+\ell_{m_i}}\nonumber\\
    &\times a^\dagger_{m_1}\ldots a^\dagger_{m_\mu}a_{a_1}\ldots a_{a_\mu}\ket{\Psi^{(0)}_v}\bra{\Psi^{(0)}_v}\nonumber\\
    &\times a^\dagger_{a_\mu}\ldots a^\dagger_{a_1}a_{m_\mu}\ldots a_{m_1}\,.
\end{align}

\subsubsection{Parity-proper cluster amplitudes}
We again begin our consideration by deriving the parity selection rule imposed on the PP amplitudes. We provide here a formal treatment which motivates and generalizes well to the case of PM amplitudes. In Sec.~\ref{Sec:ParOpSecondQuantization}, we presented the second-quantized form of  the parity operator $\Pi$.
% which affects the reversal of spatial coordinates on electronic wave functions. When acting on an many-body state in the independent-particle model, $\ket{\Psi_n}=a^\dagger_{i_1}\dots a^\dagger_{i_n}\ket{0}$, the operator $\Pi$ yields
% \begin{equation}\label{Pi_def}
%     \Pi\ket{\Psi_n}=(-1)^{\sum_n\ell_{i_n}}\ket{\Psi_n}\,,
% \end{equation}
% where $(-1)^{\ell_{i_j}}$ is the parity of the single-electron orbital $a^\dagger_{i_j}\ket{0}$ and $\ket{0}$ is the (true) vacuum state.
Using Eq.~\eqref{Eq:ParOp_2ndQuan}, it may be checked that the lowest order state $\Psi^{(0)}_v$, comprising of one valence electron above a closed-shell core, satisfies
\begin{equation}\label{cond_P}
    \Pi\ket{\Psi^{(0)}_v} =(-1)^{\ell_v}\ket{\Psi^{(0)}_v}\,,
\end{equation}
where we have again used the fact that the closed-shell core has even parity. 

Since the inter-electron Coulomb interaction is $P$-even, we require that the correlation corrections to the zeroth-order wave function  also satisfy Eq.~\eqref{cond_P}, i.e., 
\begin{equation}\label{condt_parity_PP}
    \Pi\ket{\Psi_v}=(-1)^{\ell_v}\ket{\Psi_v}\,,
\end{equation}
where $\ket{\Psi_v}=\Omega\ket{\Psi^{(0)}_v}$ is the PP equivalent of the wave function defined in Eq.~\eqref{Eq:PsivOmega}.

Expanding the wave operator $\Omega$ into singles, doubles, triples, etc., we may write the wave function $\Psi_v$ as
\begin{align}\label{B8}
    \ket{\Psi_v}&=\left(1+\sum_{ma}\rho_{ma}a^\dagger_ma_a+\frac{1}{2!}\sum_{mnab}\rho_{mnab}a^\dagger_ma^\dagger_na_ba_a\right.\nonumber\\
    &+\sum_{m\neq v}\rho_{mv}a^\dagger_ma_v+\frac{1}{2!}\sum_{mna}\rho_{mnva}a^\dagger_ma^\dagger_na_aa_v\nonumber\\
    &\left.+\frac{1}{3!}\sum_{mnrab}\rho_{mnrvab}a^\dagger_ma^\dagger_na^\dagger_ra_ba_aa_v+\ldots\right)\ket{\Psi^{(0)}_v}\,,
\end{align}
which gives
\begin{align}\label{B9}
    \Pi\ket{\Psi_v}&=(-1)^{\ell_v}\Bigg[1+\sum_{ma}(-1)^{\ell_m+\ell_a}\rho_{ma}a^\dagger_ma_a\nonumber\\
    &+\sum_{m\neq v}(-1)^{\ell_m+\ell_v}\rho_{mv}a^\dagger_ma_v\nonumber\\
    &+\frac{1}{2!}\sum_{mnab}(-1)^{\ell_m+\ell_n+\ell_a+\ell_b}\rho_{mnab}a^\dagger_ma^\dagger_na_ba_a\nonumber\\
    &+\frac{1}{2!}\sum_{mna}(-1)^{\ell_m+\ell_n+\ell_a+\ell_v}\rho_{mnva}a^\dagger_ma^\dagger_na_aa_v\nonumber\\
    &+\frac{1}{3!}\sum_{mnrab}(-1)^{\ell_m+\ell_n+\ell_r+\ell_v+\ell_a+\ell_b}\rho_{mnrvab}\nonumber\\
    &\times a^\dagger_ma^\dagger_na^\dagger_ra_ba_aa_v+\ldots\Bigg]\ket{\Psi^{(0)}_v}\,.
\end{align}

Substituting Eqs.~\eqref{B8} and \eqref{B9} into Eq.~\eqref{cond_P} and comparing like terms, we obtain the following parity selection rules for the PP cluster amplitudes
\begin{equation}\label{B91}
\begin{aligned}
    (-1)^{\ell_m+\ell_a}\rho_{ma}&= \rho_{ma}\,,\\
    (-1)^{\ell_m+\ell_v}\rho_{mv}&=\rho_{mv}\,,\\
    (-1)^{\ell_m+\ell_n+\ell_a+\ell_b}\rho_{mnab}&=\rho_{mnab}\,,\\
    (-1)^{\ell_m+\ell_n+\ell_v+\ell_a}\rho_{mnva}&=\rho_{mnva}\,,\\
    (-1)^{\ell_m+\ell_n+\ell_r+\ell_v+\ell_a+\ell_b}\rho_{mnrvab}&=\rho_{mnrvab}\,.
\end{aligned}
\end{equation}
The first of Eqs.~\eqref{B91} implies that if $\ell_m+\ell_a$ is odd then $\rho_{ma}=0$. Similar parity selection rules follow from the rest of Eqs.~\eqref{B91}.

\subsubsection{Parity-mixed cluster amplitudes}
We have shown that the use of the parity operator $\Pi$ allows us to formally prove the parity selection rules imposed on the PP amplitudes. To extend this formalism to the case of PM amplitudes, we first consider the effect of $\Pi$ on a PM single-electron orbital $\psi'_i$. Since the PM orbital $\psi'_i$ splits into two components with opposite parities, Eq.~\eqref{PNC_DHF_para}, we have
\begin{align}\label{Effect_of_parity}
    \Pi\ket{\psi'_i}&=(-1)^{\ell_i}\ket{\psi_i}+(-1)^{\ell_{\bar i}} \im\eta\ket{\bar{\psi}_i}\nonumber\\
    &=(-1)^{\ell_i}\left(\ket{\psi_i}- \im\eta\ket{\bar{\psi}_i}\right)\nonumber\\
    &=(-1)^{\ell_i}P_\eta\ket{\psi'_i}\,,
\end{align}
where we have introduced the operator $P_\eta$ which changes $\eta$ to $-\eta$.
We see that the action of $\Pi$ on $\psi'_i$ is equivalent to multiplying $\psi'_i$ with its nominal parity $(-1)^{\ell_i}$ and applying the operator $P_\eta$. Note that since changing the sign of $\eta$ does not affect terms $O(\eta^2)$, Eq.~\eqref{Effect_of_parity} is correct up to $O(\eta^3)$.

To find the PM equivalence of Eq.~\eqref{cond_P}, we first expand the PM creation operator $a'^\dagger_i$ in terms of its PP counterparts
\begin{equation}
    a'^\dagger_i = a^\dagger_i + \im\eta\sum_{\bar{j}}\gamma_{i\bar{j}}a^\dagger_{\bar{j}}\,,
\end{equation}
where we have used Eqs.~\eqref{psi'_expansion} and \eqref{restrict_coeff} (see, for example, Ref.~\cite{cohen2019quantum} for a detailed discussion on the effects of basis rotation on the second-quantization operators). As a result, up to $O(\eta)$, we have
\begin{align}
    \ket{\Psi'^{(0)}_v}&=a'^\dagger_va'^\dagger_{a_1}\dots a'^\dagger_{a_{n-1}}\ket{0}=a^\dagger_va^\dagger_{a_1}\dots a^\dagger_{a_{n-1}}\ket{0}\nonumber \\
    &+ \im\eta\left(\sum_{\bar{j}}\gamma_{v\bar{j}}a^\dagger_{\bar{j}}\right)a^\dagger_{a_1}\dots a^\dagger_{a_{n-1}}\ket{0}\nonumber \\
    &+\im\eta a^\dagger_v\left(\sum_{\bar{j}}\gamma_{a_1\bar{j}}a^\dagger_{\bar{j}}\right)\dots a^\dagger_{a_{n-1}}\ket{0}\nonumber\\
    &+\dots+\im\eta a^\dagger_va^\dagger_{a_1}\dots \left(\sum_{\bar{j}}\gamma_{a_{n-1}\bar{j}}a^\dagger_{\bar{j}}\right)\ket{0}\,,
\end{align}
which makes it clear that
\begin{equation}\label{parity_PM_Psi_0}
    \Pi\ket{\Psi'^{(0)}_v} = (-1)^{\ell_v}P_\eta\ket{\Psi'^{(0)}_v}\,.
\end{equation}
Again, we have used the fact that the nominal parity of a closed-shell core is even to remove the factor $(-1)^{\sum_a \ell_a}$ in Eq.~\eqref{parity_PM_Psi_0}. Again, we see that the action of $\Pi$ on $\Psi'^{(0)}_v$ is equivalent to multiplying $\Psi'^{(0)}_v$ with its nominal parity factor and changing $\eta\rightarrow-\eta$, where the nominal parity of the many-electron state $\Psi'^{(0)}_v$ is defined as that of its valence orbital (the closed-shell core has even nominal parity).

We now consider the effect of correlations on the nominal parity of the zeroth-order state. Although the operator $V'_c$ in Eq.~\eqref{Hamiltonian'}, which comprises of the inter-electron Coulomb interaction and the PNC-DHF potential, is not $P$-even, it is predominantly so. In other words, we may write 
\begin{equation}
    V'_c = V^{\rm PC}_c + \im\eta V_c^{\rm PNC}
\end{equation}
where $V_c^{\rm PC}$ and $V_c^{\rm PNC}$ are respectively its parity conserving and parity non-conserving parts. As a result, the perturbation $V'_c$ preserves, up to $O(\eta)$, the nominal parity of the zeroth-order state. 

This may be demonstrated more rigorously if we assume, without loss of generality, that $\Psi'^{(0)}_v$ is predominantly even and write, for brevity, $\Psi'^{(0)}_v={\rm e}^{(0)}+\im\eta{\rm o}^{(0)}$, where the letter ``e'' denotes a $P$-even wave function and the letter ``o'' denotes a $P$-odd wave function. To the first order in $V'_c$, the correlation correction to $\Psi'^{(0)}_v$ has the structure
\begin{align}\label{even_odd}
        \ket{\delta\Psi'_v} &\sim \bra{{\rm e}-\im\eta{\rm o}}V'_c\ket{{\rm e}^{(0)}+\im\eta{\rm o}^{(0)}}\ket{{\rm e}-\im\eta{\rm o}}\nonumber\\
        &+ \bra{{\rm o}-\im\eta{\rm e}}V'_c\ket{{\rm e}^{(0)}+\im\eta{\rm o}^{(0)}}\ket{{\rm o}-\im\eta{\rm e}}\,,
    \end{align}
where, for brevity, we have dropped the energy denominator and the summation over intermediate states. Expanding Eq.~\eqref{even_odd} and keeping only terms up to $O(\eta)$, we have
\begin{align}
    \ket{\delta\Psi'_v} &\sim\bra{{\rm e}}V_c^{\rm PC}\ket{{\rm e}^{(0)}}\ket{\rm e}+\im\eta\bra{{\rm o}}V_c^{\rm PNC}\ket{{\rm e}^{(0)}}\ket{\rm o}\nonumber\\
    &+\im\eta\bra{{\rm e}}V_c^{\rm PNC}\ket{{\rm o}^{(0)}}\ket{\rm o}\,,
\end{align}
which clearly shows that, parity-wise, $\delta\Psi'_v$ and thus $\Psi'_v$ have the same structure as $\Psi'^{(0)}_v$. More explicitly, we require that 
\begin{equation}\label{parity_PM_Psi'}
    \Pi\ket{\Psi'_v} = (-1)^{\ell_v}P_\eta\ket{\Psi'_v}\,.
\end{equation}

We may now use Eq.~\eqref{parity_PM_Psi'} to derive selection rules on the PM cluster amplitudes similar for those in Eqs.~\eqref{B91}. For this purpose, we first note the effect of $P_\eta$ on the $\rho'$. Since the the weak interaction introduces imaginary components to the cluster amplitudes, namely, $\rho'_{ma} = \rho_{ma}+i\eta\rho''_{ma}$ and so on, we see that changing $\eta\rightarrow-\eta$ is the same as taking the complex conjugates of these amplitudes, i.e.,
\begin{equation}\label{B19_cc}
\begin{aligned}
    P_\eta\rho'_{ma} &= (\rho'_{ma})^*\,,\\
    P_\eta\rho'_{mv} &= (\rho'_{mv})^*\,,\\
    P_\eta\rho'_{mnab} &= (\rho'_{mnab})^*\,,\\
    P_\eta\rho'_{mnva} &= (\rho'_{mnva})^*\,,\\
    P_\eta\rho'_{mnrvab} &= (\rho'_{mnrvab})^*\,.
\end{aligned}
\end{equation}

Using Eq.~\eqref{B19_cc}, we may now write Eq.~\eqref{parity_PM_Psi'} in a more explicit form. Remembering that $\ket{\Psi'_v}=\Omega'\ket{\Psi'^{(0)}_v}$, we may expand the wave operator $\Omega'$ into singles, doubles, and triples, obtaining
\begin{align}\label{B18}
    \Pi\ket{\Psi'_v}&=(-1)^{\ell_v}\left[1+\sum_{ma}(-1)^{\ell_m+\ell_a}\rho'_{ma}P_\eta a'^\dagger_ma'_a\right.\nonumber\\
    &\left.+\sum_{m\neq v}(-1)^{\ell_m+\ell_v}\rho'_{mv}P_\eta a'^\dagger_ma'_v\right.\nonumber\\
    &\left.+\frac{1}{2!}\sum_{mnab}(-1)^{\ell_m+\ell_n+\ell_a+\ell_b}\rho'_{mnab}P_\eta a'^\dagger_ma'^\dagger_na'_ba'_a\right.\nonumber\\
    &\left.+\frac{1}{2!}\sum_{mna}(-1)^{\ell_m+\ell_n+\ell_a+\ell_v}\rho'_{mnva}P_\eta a'^\dagger_ma'^\dagger_na'_aa'_v\right.\nonumber\\
    &\left.+\frac{1}{3!}\sum_{mnrvab}(-1)^{\ell_m+\ell_n+\ell_r+\ell_v+\ell_a+\ell_b}\rho'_{mnrvab}\right.\nonumber\\
    &\times P_\eta a'^\dagger_ma'^\dagger_na'^\dagger_ra'_ba'_aa'_v+\ldots\Bigg]\ket{\Psi^{(0)}_v}\,,
\end{align}
and 
\begin{align}\label{B31}
    P_\eta\ket{\Psi'_v}&=\left[1+\sum_{ma}(\rho'_{ma})^*P_\eta a'^\dagger_ma'_a\right.\nonumber\\
    &\left.+\sum_{m\neq v}(\rho'_{mv})^*P_\eta a'^\dagger_ma'_v\right.\nonumber\\
    &\left.+\frac{1}{2!}\sum_{mnab}(\rho'_{mnab})^*P_\eta a'^\dagger_ma'^\dagger_na'_ba'_a\right.\nonumber\\
    &+\frac{1}{2!}\sum_{mna}(\rho'_{mnva})^*P_\eta a'^\dagger_ma'^\dagger_na'_aa'_v\nonumber\\
    &+\frac{1}{3!}\sum_{mnrvab}(\rho'_{mnrvab})^*\nonumber\\
    &\times P_\eta a'^\dagger_ma'^\dagger_na'^\dagger_ra'_ba'_aa'_v+\ldots\Bigg]\ket{\Psi^{(0)}_v}\,.
\end{align}

Substituting Eqs.~\eqref{B18} and~\eqref{B31} into Eq.~\eqref{parity_PM_Psi'} and comparing like terms, we obtain the following selection rules for the PM cluster amplitudes
\begin{equation}\label{B19}
\begin{aligned}
    (-1)^{\ell_m+\ell_a}\rho'_{ma}&= (\rho'_{ma})^*\,,\\
    (-1)^{\ell_m+\ell_v}\rho'_{mv}&=(\rho'_{mv})^*\,,\\
    (-1)^{\ell_m+\ell_n+\ell_a+\ell_b}\rho'_{mnab}&=(\rho'_{mnab})^*\,,\\
    (-1)^{\ell_m+\ell_n+\ell_v+\ell_a}\rho'_{mnva}&=(\rho'_{mnva})^*\,,\\
    (-1)^{\ell_m+\ell_n+\ell_r+\ell_v+\ell_a+\ell_b}\rho'_{mnrvab}&=(\rho'_{mnrvab})^*\,.
\end{aligned}
\end{equation}
Clearly, if $\psi'_m$ and $\psi'_a$ have the same nominal parity then the first of Eqs.~\eqref{B19} implies that $\rho'_{ma}$ is real. On the other hand, if $\psi'_m$ and $\psi'_a$ have opposite nominal parities then this equation implies that $\rho'_{ma}$ is imaginary. The rest of Eqs.~\eqref{B19} have the same interpretation for $\rho'_{mv}$ and other double and valence triple amplitudes.

%We end this proof by stressing that although we stop with valence triples, the formalism presented here is applicable to higher cluster amplitudes as well. As a result, it is generally true that an arbitrary PM cluster amplitude is purely real if the total nominal parities of its constituent single-electron PM orbital is even whereas if this total nominal parity is odd then the PM cluster amplitude is purely imaginary.

\bibliographystyle{apsrev4-2}
\bibliography{Lit}
%\bibliography{library-apd}

\end{document}